\documentclass[pra,superscriptaddress,twocolumn]{revtex4-2}

\usepackage{amsmath}
\usepackage{physics}
\usepackage{graphicx}
\usepackage{float}
\usepackage[colorlinks=true,linkcolor=blue,citecolor=red,urlcolor=magenta]{hyperref}
\usepackage{charter}
\newtheorem{theorem}{Theorem}
\newtheorem{observation}[theorem]{Observation}

\begin{document}

\begin{abstract}
    Long-range interacting quantum systems exhibit interesting phenomena beyond the nearest-neighbours regime. Such interactions can reach further separated particles and change the behaviour of many-body correlations. Here, we analyse their structure in the antiferromagnetic long-range interaction transverse field Ising model with power-law interactions decaying as $1/r^{\alpha}$. We examine the spatial behaviour of basis-independent total multipartite correlations, called sector lengths $S_k$, of its ground state in the vicinity of critical points for different interaction exponents $\alpha$ and fixed system size. These correlation functions were shown to indicate entanglement and the existence of local realistic models, allowing us to make claims about the Bell nonlocality and entanglement in the system. We further investigate them in the context of non-stabiliserness, often referred to as quantum magic, which holds information on their classical simulability and usefulness for quantum computation. We evaluate R{\'e}nyi stabiliser entropy near and outside the criticality and show that cumulative magic, up to the critical point, increases with $\alpha$ for all of the examined system sizes. Our results provide new insights into the interplay between the long-range interactions, observable non-classical phenomena and their classical simulability.
\end{abstract}

\title{Higher-order Spatially Separated Correlations and Quantum Magic in the Long-range Transverse Field Ising Model}

\author{Pawe{\l} Cie\'sli\'nski}
\email{pawel.cieslinski@nus.edu.sg}
\affiliation{Centre for Quantum Technologies, National University of Singapore, Singapore 117543, Singapore}

\author{Wies{\l}aw Laskowski}

\affiliation{Institute of Theoretical Physics and Astrophysics, University of Gdańsk, 80-308 Gda\'nsk, Poland}

\maketitle

\section{Introduction}

Quantum entanglement and Bell nonlocality manifest themselves through correlations between measurements on individual subsystems, constituting fundamental features of quantum theory with profound implications, both foundational and practical, for modern physics and quantum technologies~\cite{Horodecki_review, Guhne_2009, Amico_review, Brunner_review}. These include quantum communication \cite{Ekert_1991, Mayers_1998, Barrett_2005, Acin_2007, Pironio_2009}, computation \cite{Linden_2001, Jozsa_2003, Harrow_2017, Bauer_2020}, and metrology \cite{Giovannetti_2004,Giovannetti_2011, Wiseman_2009, Toth_2014, Montenegro_2025}, beyond others. Another non-classical resource closely linked with observable correlations and recognised for its role in quantum computing is the quantum magic, also referred to as non-stabiliserness~\cite{Gottesman_1997, Gottesman_1998, Aaronson_2004}. It emerged from the Gottesman–Knill theorem, stating that any circuit starting from a $|0\rangle^{\otimes N}$ state and consisting only of Clifford operations (e.g. Pauli unitaries) can be efficiently simulated on a classical computer~\cite{Gottesman_1997, Gottesman_1998}. Thus, the amount of quantum magic is treated as a resource in quantum computing. The more magic the state contains, the harder it is believed to simulate it classically. 
\begin{figure}[h!]
    \centering
    \includegraphics[width=0.48\textwidth]{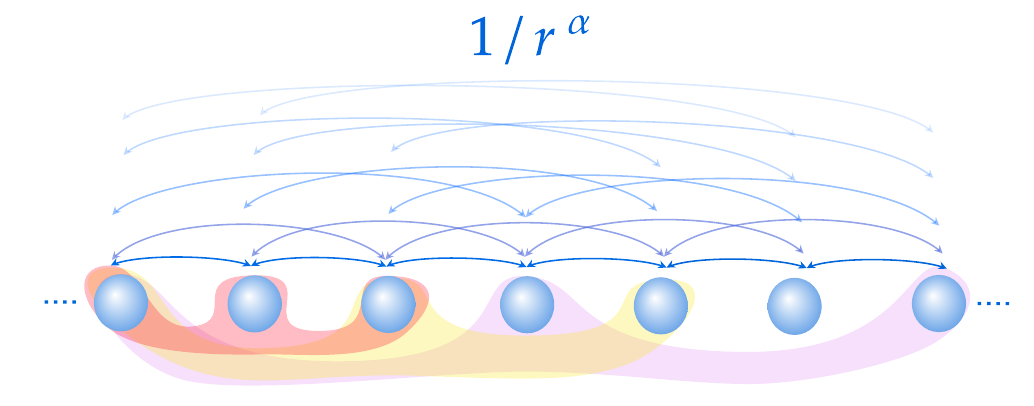}
    \caption{Higher-order correlations in the long-range transverse field Ising model (LITF). The figure above represents the main premise behind this work. Given a LITF we determine the ground state of the system for different interaction exponents $\alpha$ and calculate the basis-independent many-body correlations for evenly spaced particle configurations. The figure shows an example of the three-body case with different configurations shown in distinct colors. Based on these results, we discuss their spatial behaviour, entanglement and Bell non-local properties. Later, we move towards determining the interplay between the interaction exponent and quantum magic in different regions of the phase diagram, showing which systems are harder to simulate classically.}
    \label{fig:idea}
\end{figure}
While generating quantum states and correlations remains a central research challenge in quantum information, understanding their nature in interacting systems gives us new perspectives and broadens our understanding of the limitations on their existence and observation, see e.g. \cite{Dowling_2004, Guhne_2005, Hofmann_2014, Soldati_2021, Su_2022, Cieslinski_2023, Cieslinski_2024_fisher, Consiglio_2025, Wiesniak_2025}. Condensed matter and many-body models provide a useful platform for such investigations. This includes entanglement detection~\cite{Dowling_2004, Guhne_2005}, thermal robustness~\cite{Markham_2008, Nakata_2009}, phase transitions indicated by quantum correlation measures~\cite{Anders_2007, Hofmann_2014, Li_2024} or quantifying the amount of quantum magic and detecting phase transitions with it~\cite{Liu_2022, Fu_2022, Huang_2023, Tarabunga_2024}.  

Long-range interacting quantum models~\cite{longrange_review} are perfect candidates for such investigations. Not only are they less understood than the nearest-neighbour ones, but they also include correlation effects which do not exist in the short-range models~\cite{Lahaye_2009, Koffel_2012, Knap_2013, Vodola_2015}. Intuitively, one may also think that the longer the range of interactions, the more correlated the spatially separated particles should be. Often, due to their non-integrability, higher-order correlations that are interesting from an informational perspective are even more mysterious. Moreover, recent advancements in their exploration and new experimental capabilities have caused them to attract more attention in both condensed matter and many-body physics, as well as quantum information~\cite{longrange_review}. One such model that is especially interesting is the long-range transverse field Ising model (LITF), with power-law interactions, i.e. with coupling strength decaying as $\sim 1/r^{\alpha}$. Here we focus specifically on this model.

Higher-order multipartite correlations allow one to detect and characterise entanglement~\cite{Horodecki_review}, non-locality~\cite{Brunner_review} and quantum magic~\cite{Leone_2022}. One insightful way to quantify them, which also occurs in the error correction research, is by using the sector lengths~\cite{Coffman_2000}. They capture the total correlations of a given size, and importantly, are basis independent. This means that they can unify, e.g. the notions of a standard as well as the staggered magnetisation in the one-body correlation case. For higher orders, they form entanglement~\cite{Hassan_2008, Badziag_2008, Tran_2015, Ketterer_2019, Wyderka_2020} and non-locality~\cite{Zukowski_2002} criteria that are especially feasible experimentally and can be measured with randomised measurements~\cite{Cieslinski_2024}.

In this work, we explore the spatial structure and behaviour of many-body correlations through the sector lengths near the critical point in the ground state of a one-dimensional antiferromagnetic LITF and quantum magic dependence on the interaction exponent $\alpha$ near and outside of the criticality using tensor network calculations based on the density matrix renormalisation group (DMRG), complemented with exact diagonalisation. See Fig.~\ref{fig:idea} for its graphical representation. We observe that these basis-independent correlations for a fixed system size scale algebraically near the critical point for evenly spaced sets of multiple particles and show that they allow one to infer entanglement present in the system. We discuss their structure and strength for changing interaction exponents and show that, up to seven-body correlations, there exists a local realistic description of the observed statistics under two-settings correlation Bell inequalities. Furthermore, we examine the dependence of quantum magic on $\alpha$ and calculate it explicitly for the extreme cases. Our results show that non-stabiliserness of the LITF ground state is highly influenced by the pace of interaction decay. More specifically, we find that the cumulative magic, up to the critical point and for the examined system sizes, grows with $\alpha$ -- the faster the couplings decay, the higher the total magic in the state over the interaction strength parameter is. Our findings suggest that some of the long-range interacting systems contain more resources from the quantum computing perspective and provide valuable insights into their observable correlation structure.

\section{Physical setting and preliminaries}
\subsection{Long-range power-law interaction transverse field Ising model}
We start by introducing the physical setting of our interest. We chose to focus on the extended version of the well-studied one-dimensional transverse field Ising model that captures interactions beyond the nearest neighbours. If one chooses the couplings to decay as a power of the distance separating the spins-$1/2$ particles, the LITF Hamiltonian is given as
\begin{equation}
    H=\frac{\lambda}{2}\sum_{i \neq j}^L \frac{1}{|i-j|^{\alpha}} \sigma^x_i \sigma^x_j+ \sum_i^L \sigma^z_i,
    \label{eq:Hamiltonian}
\end{equation}
where $\sigma^x_i, \sigma^z_j$ are the Pauli matrices acting non-trivially on the sites $i$ and $j$, $\alpha \geq 0$ is the interaction exponent and $\lambda$ is the rescaled coupling parameter. Additionally, we constrain ourselves to the antiferromagnetic regime with $\lambda>0.$ This model was studied in the past from both theoretical and experimental~\cite{Koffel_2012, Knap_2013, Vodola_2015, Fey_2016, Britton_2012, Islam_2013, Bohnet_2016, Yang_2019} perspectives. In this work, our main focus will be on the behaviour and structure of the many-body correlations in the ground state of the above Hamiltonian. Notably, even the two-body correlations were shown to have non-trivial features in the past. It was shown that in the $z$ polarised phase, even when it is gapped, two-body connected correlation functions $\langle \sigma_{L/2} \sigma_{L/2+r} \rangle-\langle \sigma_{L/2}\rangle \langle \sigma_{L/2+r} \rangle$ in the $x$ and $z$ plane decay as $1/r^{\sim \alpha}$~\cite{Koffel_2012}. In \cite{Vodola_2015} it was further observed that they can decay with a hybrid behaviour that is exponential at short distances and algebraic at long ones. Specifically, in the antiferromagnetic phase and $\alpha >1$, a clear algebraic behaviour at long distances was reported. Similar findings were also presented for the XXZ model with dipole ($\alpha=3$) interactions~\cite{Hauke2010, Peter_2012}. For the full correlations near criticality, the algebraic decay is expected, and in the antiferromagnetic and paramagnetic phases, the corresponding order persists for all $\alpha$~\cite{Vodola_2015}. 

\subsection{Sector lengths as basis independent correlation functions}

A standard approach to studying the correlation functions in many-body spin-$1/2$ (or qubit) systems consists of examining an expectation value of a tensor product of Pauli matrices between particles $i$ and $j$, i.e. $\langle \sigma_i \sigma_j\rangle$. However, such quantities are intrinsically a basis-dependent notion, which can be problematic if the state under consideration is unknown. In general, it would be beneficial to have a single quantity that captures correlations in any basis and provides additional information on entanglement or Bell non-locality. Note that basis independence is what distinguishes classical correlations from the ones observed in quantum entangled states. Sector lengths~\cite{Coffman_2000}, also known as Shor-Laflamme, or weight, enumerators in error correction research~\cite{Shor_1997}, and corresponding correlation tensor-based criteria are the objects that encompass it all. Any $L$-qubit state can be expressed in a basis of the tensor product of Pauli matrices basis as
\begin{eqnarray}
    \rho=\frac{1}{2^L} \sum_{\mu_1,\mu_2, \cdots , \mu_L=0}^3 T_{\mu_1,\mu_2, \cdots , \mu_L} \sigma^{\mu_1}_1\cdots \sigma^{\mu_L}_L,
\end{eqnarray}
where we use the convention $(\sigma^0,\sigma^1,\sigma^2,\sigma^3)=(\openone,\sigma^x,\sigma^y,\sigma^z)$ and $T_{\mu_1,\mu_2, \cdots , \mu_L}=\mathrm{Tr}(\sigma^{\mu_1}_1\cdots \sigma^{\mu_L}_L \rho)$ are the correlation tensor $\hat{T}$ elements. The $k$-body sector length is defined as
\begin{eqnarray}
    S_k(\rho)=\sum_{\text{$k$ non-zero indices}} T_{\mu_1,\mu_2, \cdots , \mu_L}^2.
\end{eqnarray}
In the case of $k=1$ it quantifies the sum of all Bloch vectors' length in informational terms, or a basis-independent sum of squares of local magnetisations from the condensed matter perspective. The sum of all sector lengths divided by $2^L$ yields the state purity, while higher-order sector lengths define various entanglement criteria~\cite{Cieslinski_2024}. The one which we will focus on is given as~\cite{Hassan_2008, Badziag_2008, Tran_2015}
\begin{eqnarray}
    \text{if }S_L(\rho)>1 \Rightarrow \rho \text{ is entangled.}
\end{eqnarray}
Furthermore, if one limits the measurements to a chosen plane, say $(x,y)$, then if
\begin{eqnarray}
    \text{if }S_L(\rho)|_{(x,y)}<1,
\end{eqnarray}
there exists a local realistic explanation of the observed correlation function, hence no $L$-body two-setting correlation-based Bell inequality can be violated with $\rho$~\cite{Zukowski_2002}. Note that sector lengths are experimentally feasible to measure in a direct way or through the use of randomised measurements.

Given that $S_k$ have clear interpretation and can be used to infer strictly quantum properties of the state, an interesting question in light of the studied spin model arises. How does the interaction exponent $\alpha$, which controls the decay of interaction, affect the spatial behaviour of these basis-independent many-body correlations? This would be especially significant near the criticality where the reduced states' entanglement has distinct properties. For example, in the short-range interaction case, the multipartite entanglement grows exponentially~\cite{Hofmann_2014, Su_2022}. Since the standard two-body correlation functions decay algebraically with a given critical exponent depending on $\alpha$~\cite{Fisher_1972,Dutta_2001, Defenu_2017} we expect similar effects from the sector lengths. However, in the case of a multipartite $S_k$, this problem is not so straightforward. Moreover, a change in the structure of correlations $T_{\mu_1,\mu_2, \cdots , \mu_L}$ is also an interesting problem which was never studied before. Note that we are not interested in universal scaling laws, but rather the informational quantitative content of these states and their dependence on the interaction structure for a fixed system size.

Changing the correlation structure and its behaviour also affects the amount of quantum magic present in the state. Quantum magic is a resource believed to be responsible for the advantage of quantum computations over classical ones~\cite{Gottesman_1997, Gottesman_1998, Aaronson_2004}. Beyond its purely computational importance, its interplay with the correlation structure and resources such as entanglement or Bell nonlocality has recently become of interest~\cite{Howard_2012, Howard_2014, Howard_2015, Tirrito_2024, Bejan_2024, Iannotti_2025, Gu_2025, Macedo_2025, Cusumano_2025, Cieslinski_2026, Viscardi_2026}.
Quantum magic captures the degree of non-stabiliserness and can be measured with generalised sector lengths through the use of stabiliser R{\'e}nyi entropy (SRE)~\cite{Leone_2022}. The specific measure we will focus on is defined as
\begin{equation}
    M_2(|\psi\rangle)=-\log \frac{1}{2^L} \sum_{\mu_1, \ldots, \mu_L=0}^3 T_{\mu_1 \cdots \mu_L} ^4.
\end{equation}
Elements of the above sum are just modified $k$-body sector lengths where the power of the correlation tensor element is changed from $2$ to $4$. Thus, we can write
\begin{equation}
    M_2(|\psi\rangle)=-\log \frac{1}{2^L} \sum_{k=0}^{L} S_k^{(4)}.
    \label{eq:magic}
\end{equation}
It is easy to see that any stabiliser state, i.e. the state containing only perfect $\pm 1$ correlations, results in a 0 magic and any divergence from that causes its growth. In our work, we are interested in the behaviour of SRE as a function of interaction exponent $\alpha$ as this would give us an insight into the problem of simulating long-range interacting models.

\section{Results}

In order to study many-body sector length spatial scaling and quantum magic in the ground state of LITF (\ref{eq:Hamiltonian}) we first determine the critical points $\lambda_c$ for $5/3 \leq \alpha \leq 3$ with a step of $1/3$. The choice of this interval for $\alpha$ is motivated by the universality classes. For $\alpha<5/3$ the mean-field approach is exact, and for $\alpha>3$ the phase transition is within the short-range Ising model universality class~\cite{Dutta_2001}. In the studied cases, the system has non-trivial varying critical exponents, which makes it more interesting to study, see e.g. \cite{Koffel_2012, Knap_2013, Vodola_2015, Fey_2016}. In all our considerations, we are using tensor network calculations implemented with ITensor in Julia using DMRG~\cite{ITensor} complemented with exact diagonalisation results for comparison. The bound dimension $\chi$ for the studied model is usually chosen to be $\chi \leq 100$. In our calculations, we chose it such that the maximal truncation error was never greater than $10^{-16}$ and that the results agree with exact diagonalisation for small systems. In the case of $L=50,100,200$ our simulations were precise enough when $\chi \leq L/2$. However, for smaller systems for which quantum magic was examined, we went as far as  $\chi \leq 800$ for $L=25$.

\begin{figure}
    \centering
    \includegraphics[width=0.9\linewidth]{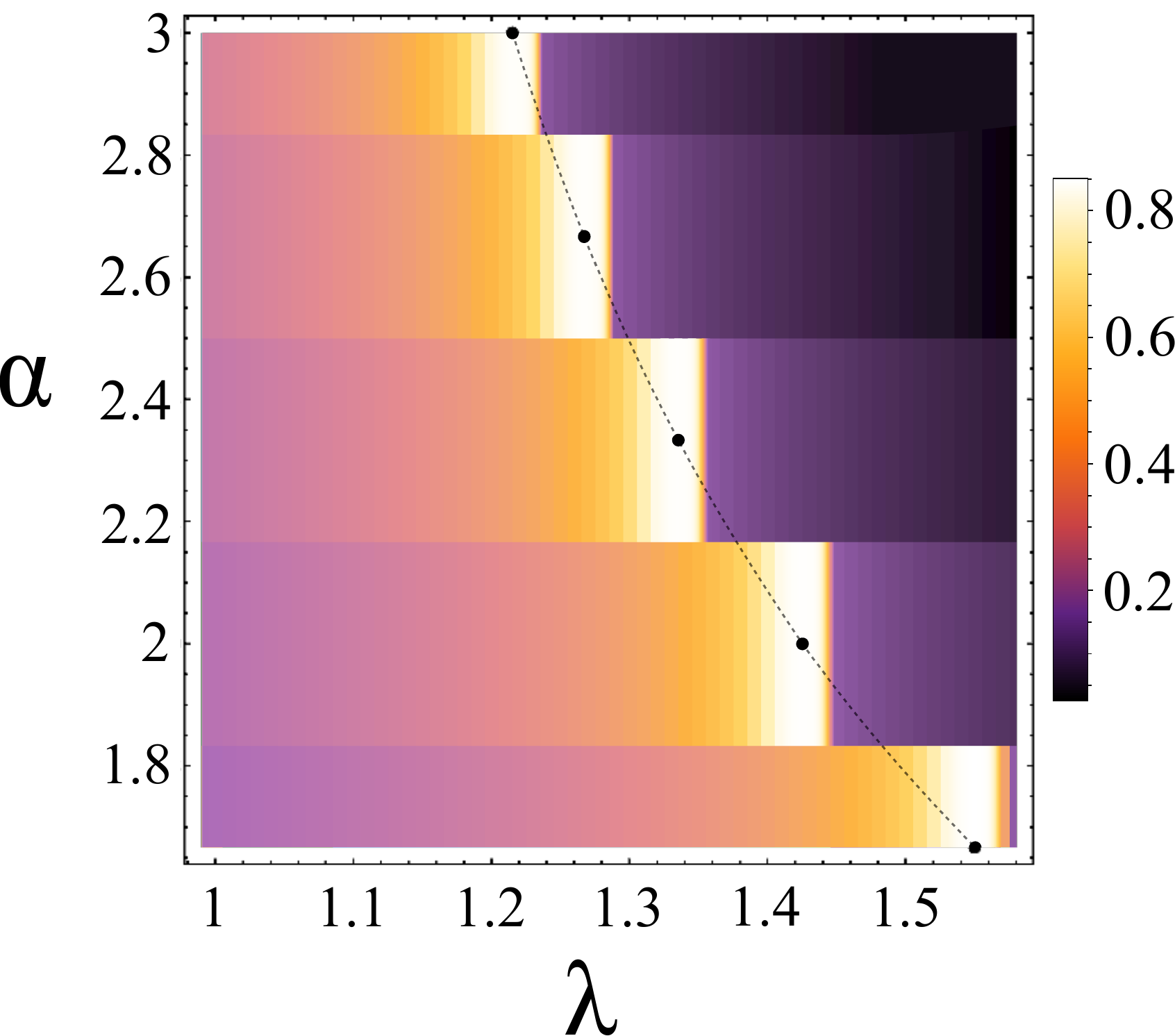}
    \caption{\textit{Half chain entanglement entropy near the critical points.} The background colours represent the value of a half chain entanglement entropy evaluated on the ground state of LITF Hamiltonian for each of the studied interaction exponents $\alpha$ and $L=200$. Its maximal value signifies the location of pseudo critical points $\lambda(L=200,\alpha)$. Analogous calculations for $L=50,100,200$ allowed for their extrapolation and estimation of $\lambda_C(\alpha)$ marked with black dots. The dashed black line interpolating these points serves as a guide for the eye.}
    \label{fig:phase_diagram}
\end{figure}

The phase diagram in the $(\lambda, \alpha)$ plane is already present in the literature, see e.g.~\cite{Koffel_2012}. Because of that, for the purpose of our study, we have only determined the pseudo critical points $\lambda_c(L, \alpha)$ for $L=50,100,200$ and extrapolated $\lambda_c(\alpha)$ for the chosen interaction exponents. This was done by evaluating the half chain entanglement entropy defined as $-\Tr \rho_{L/2}\log(\rho_{L/2})$, where $\rho_{L/2}$ is the reduced density matrix of the ground state after tracing the first $L/2$ parties. Our results agree with the previous extrapolations within a few percentage errors \cite{Koffel_2012, Knap_2013, Vodola_2015, Fey_2016}. Explicit critical points are $\lambda_c(\alpha)=\lbrace 1.53269, 1.39778, 1.31036, 1.23973, 1.18929 \rbrace$ for $\alpha= \lbrace 5/3,2,7/3,8/3,3\rbrace$ respectively. The half chain entanglement entropy in the vicinity of critical points and studied $\alpha$ is presented in Fig.~\ref{fig:phase_diagram}. Further in this section, we will focus on the multipartite sector lengths scaling with distance near criticality for $L=100$ and the behaviour of magic in a broader parameter range for smaller system sizes of $L\leq 25$.

\subsection{Multipartite correlations scaling with distance ($L=100$)}

The most commonly studied form of correlation functions are the two-point correlations, which are especially important from the perspective of condensed matter physics.  
Previous results on the LITF model have shown a surprising fact that the two-body correlations scaling can follow $\langle \sigma^\mu_i \sigma^\nu_{i+r} \rangle \sim r^{-\kappa \alpha}$ even outside of criticality, with $\kappa$ being a constant of $O(1)$ order. Near the critical point, the exponent governing this decay depends non-trivially on $\alpha$ outside of the mean-field and short-range universality class. Since two-body sector lengths are quadratic sums of such terms, they too should decay algebraically at large distances with polynomial or inverse polynomial behaviour at smaller $r$, i.e. we expect them to follow $\mathrm{poly}(1/r)$ or $\mathrm{poly}(r)^{-\gamma}$. In other regions, exponential corrections are needed~\cite{Vodola_2015}. 

In our work, we are not interested in determining the critical exponents, but rather exploring the qualitative differences between the behaviour of correlations and sector lengths for different interaction exponents $\alpha$ and supports of observables $k$ in systems of a fixed size of $L=100$. For two-point correlations, the notion of distance between qubits is straightforward; however, this becomes more arbitrary in the many-body case, as there exist many possible spatial configurations of particles which affect the quantum correlations, see e.g.~\cite{Hofmann_2014}. We choose to study the most natural case where all of the qubits are equally separated, i.e. we examine the correlations between nodes enumerated by $(L/2, L/2+r, \cdots, L/2+[k-1]r)$ (see Fig.~\ref{fig:idea}).

\begin{figure*}[ht!]
    \centering
    \includegraphics[scale=0.8]{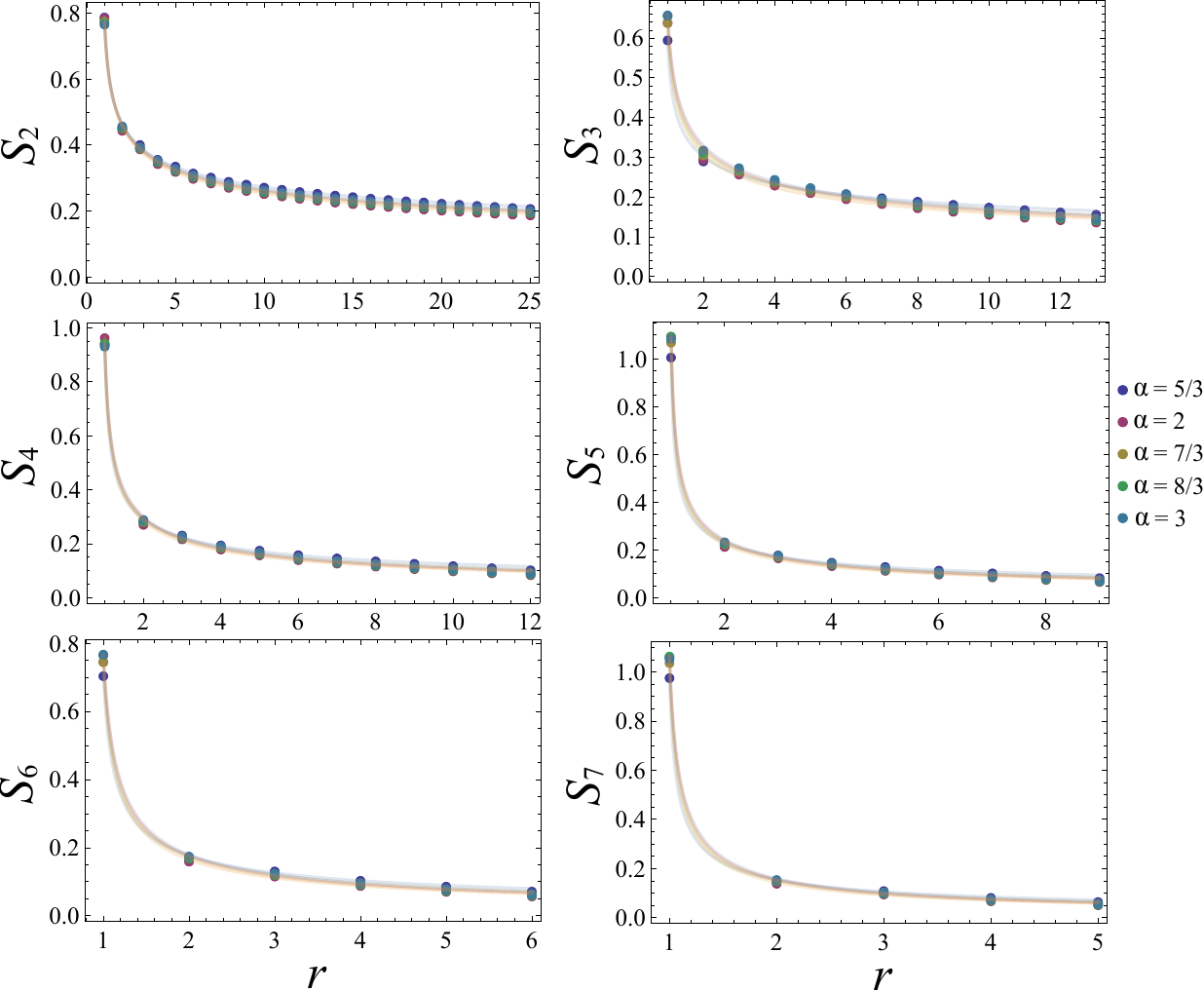}
    \caption{\textit{Spatial behaviour of multipartite correlations in the long-range interaction Ising model}. The above figure shows sector lengths $S_k$ spatial behaviour with equidistant particles in the ground state of Hamiltonian (\ref{eq:Hamiltonian}) in the vicinity of a critical point for $k\leq 7$ and $L=100$. The plotted points in the main figures are data calculated using DMRG. We find that they follow $(a+b \cdot r)^{-\gamma_k}$ with an estimated variance in the range of $10^{-5}-10^{-6}$, which for sufficiently large distances becomes algebraic. The pseudo scaling exponents $\gamma_k(\alpha)$ are not monotonous in $\alpha$. For all supports of correlation functions $k$, we observe that the pseudo scaling exponents increase and decrease interchangeably with changing $\alpha$, but in general $\alpha=5/3$ results in the smallest pace of correlations decay. For a more detailed discussion of their spatial behaviour, entanglement and Bell nonlocality, see the main text.}
    \label{fig:sector_lengths}
\end{figure*}

Our findings for the behaviour of $S_k$ for $L=100$ are shown in Fig.~\ref{fig:sector_lengths}. We find that in spite of $k$ our results follow $(a+b \cdot r)^{-\gamma}$ with estimated variances in the range of $ 10^{-5}-10^{-6}$. Note that, for large distances, it reduces to $\sim 1/r^{\gamma}$ as expected, and the scaling is algebraic. Here, large distances mean $r \gg a/b$ and in our approximation, it holds that $a/b<1$. This suggests that the total correlations, i.e. correlations independent of basis, can also manifest the standard scaling properties for $r \gg 1$. Again, such behaviour is observed for all studied sectors $S_k$ and $L=100$. An interesting conclusion that separates the standard intuitions for correlation functions from sector lengths is that different pseudo scaling exponents $\gamma_k(\alpha)$ are not monotonic in $\alpha$. This behaviour could disappear at large system sizes, but the studied $L$ is even greater than the currently available number of controlled systems in experiments~\cite{longrange_review}. Thus, their analysis is still relevant. For $k=2$ the decay is the fastest for $\alpha=2$, while for $k>2$ it happens for $\alpha=8/3$. It seems interesting that different power-law interaction powers favour second and higher-order correlations. However, as intuitively expected, we observe in general that the smallest value of the studied $\alpha=5/3$, i.e. the slowest decay of long-range interactions, results in the smallest pace of correlations decay. For all $k$ we find that the pseudo scaling exponents computed for a fixed system size of $L=100$ increase and decrease interchangeably with changing $\alpha$. At last, $\gamma_k$ increase with $k$ for $\alpha \neq 5/3$ and $8/3$. For explicit values see Table.~\ref{tab:gamma}. The above observations can be summarised as follows.

\begin{observation}
    For $L=100$, all examined $\alpha$ and $k$, the latter being the support of a basis-independent correlation function $S_k$ called sector length, in the vicinity of the critical points $\lambda_c(\alpha)$, we observe  
    \begin{equation}
        S_k(r)=(a_k+b_k \cdot r)^{-\gamma_k}
    \end{equation}
    with $r$ denoting the distance separating each of the particles.
    The fastest decay of many-body correlations ($k>2$) for occurs for
    \begin{equation}
        \gamma_k(\alpha)_{max}=\gamma_k(8/3).
    \end{equation}
     Moreover, we report that for $\alpha \neq 5/3$ and $8/3$
    \begin{equation}
        \gamma_k < \gamma_{k'}, \quad \text{for } k<k'.
    \end{equation}
\end{observation}

\begin{table}[]
\caption{\label{tab:gamma} Pseudo critical exponents for the multipartite sector lengths $S_k$ in ground state of LRIM (\ref{eq:Hamiltonian}) and $L=100$ rounded up to the third decimal place. The exponents are calculated by fitting the DMRG data according to $(a+b \cdot  r)^{-\gamma_k}$.}
\begin{tabular}{c | c c c c c}
\hline
     $\alpha$ & 5/3 & 2 & 7/3 & 8/3 & 3 \\ \hline
     $\gamma_2$ & 0.254 & 0.287 & 0.275 & 0.280 &0.275 \\
     $\gamma_3$ & 0.248 & 0.312 & 0.301 & 0.322 & 0.32 \\
     $\gamma_4$ & 0.395 & 0.395 & 0.448 & 0.471 & 0.466 \\
     $\gamma_5$ & 0.43 & 0.515 & 0.498 & 0.528 & 0.524 \\
     $\gamma_6$ & 0.511 & 0.605 & 0.585 & 0.618 & 0.613 \\
     $\gamma_7$ & 0.558 & 0.665 & 0.647 & 0.689 & 0.686 \\  \hline
\end{tabular}
\label{tab:gamma}
\end{table}

For all possible parameters under consideration, correlations of two-qubit reduced states are non-zero, up to the set precision, only on the diagonal of the correlation tensor, i.e. they contain only $T_{xx}, T_{yy}$ and $T_{zz}$ terms. However, the observed correlations are too small to detect entanglement or infer Bell nonlocality via the criteria defined in the previous section. Here, it is important to note that the one-body correlations are $\approx 0$ and the two-body states belong to the family of Bell-diagonal states~\cite{Horodecki_review}.  Nevertheless, we see that correlations between $k=5,7$ qubits allow one to detect entanglement solely on the sector length's value. This happens for the nearest neighbouring qubits as $S_k>1$ for $k=5$ and all $\alpha$ as well as $k=7$ for $\alpha>5/3$ with both maxima at $\alpha=2$. Despite the fact that five and seven-body correlations violate the separability condition, they are spread over different planes with $S_k |_{(i,j)}<1$. Thus, we conclude that there exists a hidden variable local realistic model for all of the studied correlation functions. The behaviour of the nearest neighbour sector lengths is also non-trivial with respect to $\alpha$ (plot points corresponding to $r=1)$. For $k=2$ they decay with growing $\alpha$ and grow in the case of $k=3$. In the case of $\alpha \rightarrow \infty$ the three-body reduced state is genuinely entangled for the neighbouring qubits~\cite{Hofmann_2014}. This suggests that the correlations should grow further, or they do not have a leading contribution to the genuine entanglement as in~\cite{Tran_2017}. The remaining sectors, except for $k=6$, act in a similar manner as $\gamma(\alpha)$ - they grow and decrease interchangeably, with maximum at $\alpha=2$. For $k=6$ they are approximately non-decreasing. The overall behaviour of sector lengths, in contrast to the standard correlation functions, points to the $1/r^{\gamma_k}$ decay of entanglement and nonlocality detection capabilities for $k \ll L$.

\begin{figure*}[ht!]
    \centering
    \includegraphics[width=0.95
    \linewidth]{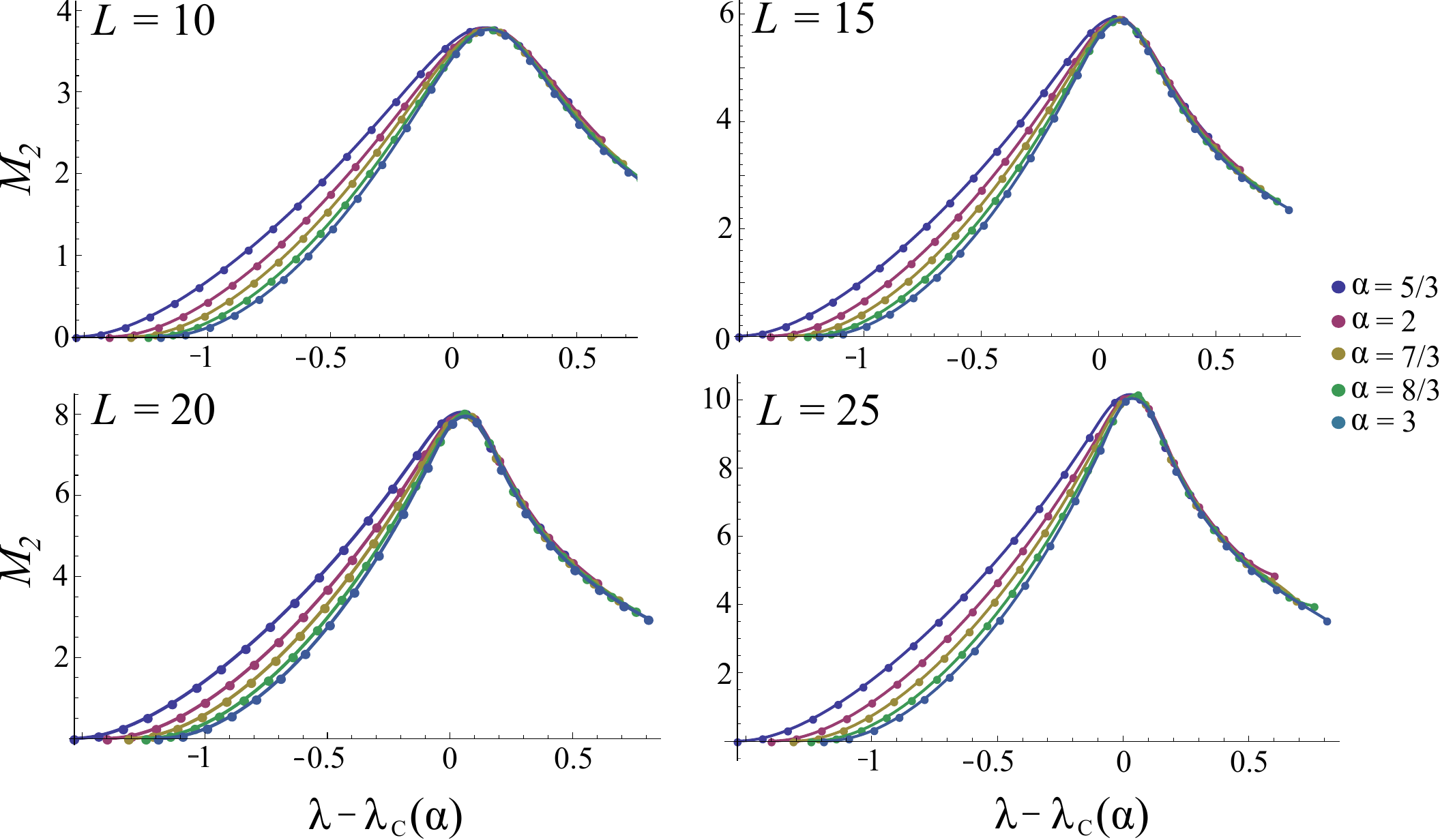}
    \caption{\textit{Ground state R{\'e}nyi stabiliser entropy $M_2$ dependence on the interaction strength $\lambda$ and exponent $\alpha$.} In the above, R{\'e}nyi stabiliser entropy $M_2$ (\ref{eq:magic}) evaluated on the ground state of LITF (\ref{eq:Hamiltonian}) is plotted over the coupling strength $\lambda$ for different values of the interaction exponent $\alpha$ and system sizes $L=10,12,20,25$. Each data point (coloured markers) was obtained using ground state MPS and a numerical technique based on the tensor cross interpolation~\cite{Kozic_2025} to evaluate the magic monotone. After rescaling the $x$-axis to $\lambda-\lambda_C(\alpha)$, where $\lambda_C(\alpha)$ are the $\alpha$ dependent critical points, the order parameter nature of SRE becomes easily visible. Moreover, using interpolation (solid coloured lines) one can evaluate the cumulative magic, see ~(\ref{eq:cum_magic}), and observe that for $\lambda\leq\lambda_C$ it increases with $\alpha$. For a more detailed discussion see the main text. }
    \label{fig:magic}
\end{figure*}

We observed that nearest neighbour qubits can be correlated in a way that allows one to detect entanglement in a very simple manner. This configuration of particles is, in fact, the one with the most visible dependence of correlations on the interaction exponent $\alpha$. The greatest differences are seen for the odd sectors. However, we find that $S_k(r)$ for distinct $\alpha$ do not differ significantly as $r$ increases. This points at the meaningful short-range correlation dependence on $\alpha$, and shows that introducing power-law decaying interactions to the system causes no qualitative change in observable correlations other than in the unseparated qubits configuration.

\subsection{Full-body sector lengths and quantum magic}

From the sector lengths results, it is clear that the studied states are not stabilisers, i.e. they do not have perfect $\pm 1$ correlations. The question we want to examine now is the effect of long-range interactions on the amount of non-stabilizerness.
In order to compute quantum magic in the ground state of (\ref{eq:Hamiltonian}) using stabiliser R{\'e}nyi entropy (\ref{eq:magic}), one has to evaluate all $2^L$ correlations. As the number of correlation tensor elements grows exponentially with the number of qubits, we limit the magic computations to the system of $L\leq25$. Recently, new advanced techniques for magic evaluation in more than a few-qubit systems were proposed~\cite{Huang_2023, Kozic_2025}. In our calculations, we use one of such recently proposed techniques based on the tensor cross interpolation~\cite{Kozic_2025} applied to the ground states obtained in the previous section. 
For the smallest system size ($L=10$), we have checked the accuracy of the obtained results with exact diagonalisation and straightforward DMRG calculations for $\alpha=5/3$. By straightforward DMRG, we refer to computing all of the Pauli strings directly from the ground state MPS and then evaluating SRE. Relative errors between ED and straightforward DMRG were $10^{-6}-10^{-7}$. For the technique from \cite{Kozic_2025}, a constant offset as compared to ED was observed. After rescaling, the highest relative error was given by $\approx 3.65 \cdot 10^{-6}$.

\begin{figure}[h!]
    \centering
    \includegraphics[width=0.95\linewidth]{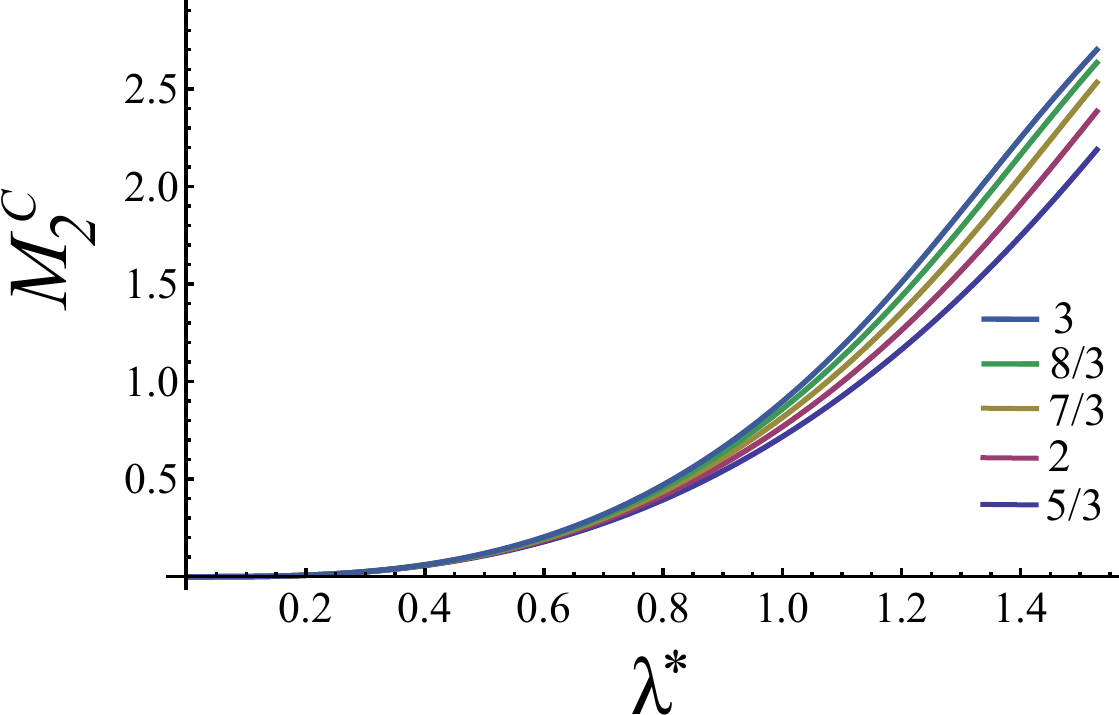}
    \caption{\textit{Cumulative magic $M_2^C$ as a function $\lambda^{*}$ for $L=10$.} Each colour represents a different value of the interaction exponent $\alpha$ and $\lambda^*$ stands for the upper integration limit of SRE. At each point, below the critical $\lambda_C$, the greater the $\lambda$, the greater the value of the cumulative magic is.}
    \label{fig:cumulative_magic}
\end{figure}

SRE as a function of $\lambda$ for different $\alpha$ and $L=10,15,20,25$ is presented in Fig.~\ref{fig:magic}. For each $\alpha$,  as expected, the maximum of $M_2$ magic is obtained at the critical point $\lambda_C$. Before reaching that point, each curve (despite the $x$-axis shift) grows at a different pace depending on $\alpha$.   
Due to different $\lambda_c(\alpha)$ for each $\alpha$, it is hard to compare the amount of magic for a single choice of parameter $\lambda$ in a quantitative way. However, this becomes possible by looking at the cumulative magic 
\begin{eqnarray}
    M^C_2(|\psi \rangle, \lambda^{*} )=\int_0^{\lambda^{*}} M_2(|\psi \rangle) d\lambda,
\end{eqnarray}
which captures the total amount of quantum magic for all $\lambda \leq \lambda^{*}$. By analysing the obtained data and the interpolating functions, we make the observation below. 
\begin{observation}
    For all of the studied system sizes $L$ and interaction exponents $\alpha$, the cumulative quantum magic measured by $M_2$ (\ref{eq:magic}) 
    is ordered with respect to $\alpha$
    \begin{equation}
        M^C_2(\alpha,\lambda^{*})\leq M^C_2(\alpha',\lambda^{*}), \quad \text{ for }  \alpha < \alpha'.
        \label{eq:cum_magic}
    \end{equation}
    and all $\lambda^{*}\leq\max_{\alpha} \lambda_C(\alpha)=\lambda_C(5/3)$.
\end{observation}
An example of this observation showing the cumulative magic as a function of $\lambda^{*}$ for $L=10$ is provided in Fig.~\ref{fig:cumulative_magic}. 
After crossing the critical point, all of the curves in Fig.~\ref{fig:magic} decrease in a similar way, with bigger deviations for the highest values of $\lambda$ for $L=25$. This is, however, most likely due to the numerical errors, as very high bond dimensions ($\chi \geq 800$) were required to reach the ground state energy convergence.

In addition to the above results, we examine the ground state of (\ref{eq:Hamiltonian}) in the limit of large coupling $\lambda \gg 1$. The approximated LITF Hamiltonian in this regime becomes 
\begin{equation}
    H_{{\rm eff}} \sim \sum_{i \neq j} \frac{1}{|i-j|^{\alpha}} \sigma^x_i \sigma^x_j,
\end{equation}
and its ground state for $\alpha\neq0$ is given by the equal superposition of two N{\'e}el states in the $\sigma^x$ basis~\cite{Koffel_2012}. However, the ground state subspace is two-fold degenerate due to commutativity with $P_z=\Pi_i \sigma_i^z$, and from an informational perspective, the analysis of non-stabiliserness of any state in the subspace is interesting. With suitable parametrisation, one can define its representative as
\begin{equation} \label{eq:ground_state_rep}
    |\psi_0\rangle_{{\rm eff}}=\cos(\beta) |+-+ \cdots\rangle + \sin(\beta) |-+- \cdots \rangle,
\end{equation}
where $|\pm\rangle$ are $\pm1$ eigenvectors of $\sigma^x$. In order to compute the considered magic monotones, one has to evaluate the full correlation tensor $\hat{T}$. Up to a change of basis, this state is equivalent to the nonsymmetric GHZ state. Let $\pi(\cdot)$ represent all permutations of a string of length $L$ and $i^n$ stand for $n$ copies of element $i$, e.g. $z^3=zzz$. Then the non-zero correlation tensor elements are
\begin{eqnarray*}
&&T_{\pi(x^n 0^{L-n})}=\pm \cos(2\beta), \text{for odd $n$} \\
&&T_{\pi(x^n0^{L-n})}=\pm 1, \text{for even $n$}\\
&&T_{\pi(y^n z^{L-n})}=\pm \sin(2\beta), \text{for even $n$}
\end{eqnarray*}
From the above, we get
\begin{equation}
    \sum_{k=0}^{L}S_k^{(4)}=2^{L-3}[7+\cos(8\beta)].
\end{equation}
Finally, the SRE for the ground state representative (\ref{eq:ground_state_rep}) in the small magnetic field approximation is given as
\begin{eqnarray}
    M_2(|\psi \rangle_{{\rm eff}})=-\log \left(\frac{1}{8}[7+\cos(8\beta)] \right).
\end{eqnarray}
For N{\' e}el states and their equal superposition ($\beta=\pi/4$) quantum magic vanishes, as should be for product of $\sigma_x$ eigenstates and GHZ states. For any other $\beta$ non-stabiliserness is non zero, reaching its maxima of $\log (4/3)$ for $\beta=\pi/8$. 

\section{Conclusions}

In this work, we have examined the behaviour of quantum correlations in the ground state of the long-range interaction Ising model with power-law decay. We focused on the spatial behaviour of basis-independent multipartite correlation functions $S_k$ in the vicinity of critical points for different interaction decay exponents in a system of 100 qubits. We found that these functions decay algebraically with equidistant separation between particles for all $k$ and examined their ability to witness entanglement and Bell nonlocality. In fact, all of the studied correlation functions admit a local hidden variable model for two-setting correlation scenarios and can reveal entanglement only when the particles are not separated. Then, we directed our attention to the role of the discussed exponent $\alpha$ on the amount of quantum magic present in the ground states of the studied Hamiltonian for a broad range of coupling strength $\lambda$. We found that, within the common pre-critical range, the cumulative stabiliser R{\'e}nyi entropy increases with $\alpha$ for all of the examined cases, providing new insight into the quantum magic resources present in long-range interacting systems and shedding new light on their classical simulability. 

\section*{Acknowledgements}
We thank Ankit Kumar, Felix Huber, Marcin Wieśniak and William E. Salazar for fruitful discussions on the DMRG calculations and results analysis.  The authors acknowledge the support of the Polish National Science Centre (NCN) within
the Preludium Bis project (Grant No. 2021/43/O/ST2/02679). 
This project is supported by the National Research Foundation, Singapore through the National Quantum Office, hosted in A*STAR, under its Centre for Quantum Technologies Funding Initiative (S24Q2d0009). 

\bibliographystyle{apsrev4-2}
\bibliography{ref}

\begin{thebibliography}{77}%
\makeatletter
\providecommand \@ifxundefined [1]{%
 \@ifx{#1\undefined}
}%
\providecommand \@ifnum [1]{%
 \ifnum #1\expandafter \@firstoftwo
 \else \expandafter \@secondoftwo
 \fi
}%
\providecommand \@ifx [1]{%
 \ifx #1\expandafter \@firstoftwo
 \else \expandafter \@secondoftwo
 \fi
}%
\providecommand \natexlab [1]{#1}%
\providecommand \enquote  [1]{``#1''}%
\providecommand \bibnamefont  [1]{#1}%
\providecommand \bibfnamefont [1]{#1}%
\providecommand \citenamefont [1]{#1}%
\providecommand \href@noop [0]{\@secondoftwo}%
\providecommand \href [0]{\begingroup \@sanitize@url \@href}%
\providecommand \@href[1]{\@@startlink{#1}\@@href}%
\providecommand \@@href[1]{\endgroup#1\@@endlink}%
\providecommand \@sanitize@url [0]{\catcode `\\12\catcode `\$12\catcode `\&12\catcode `\#12\catcode `\^12\catcode `\_12\catcode `\%12\relax}%
\providecommand \@@startlink[1]{}%
\providecommand \@@endlink[0]{}%
\providecommand \url  [0]{\begingroup\@sanitize@url \@url }%
\providecommand \@url [1]{\endgroup\@href {#1}{\urlprefix }}%
\providecommand \urlprefix  [0]{URL }%
\providecommand \Eprint [0]{\href }%
\providecommand \doibase [0]{https://doi.org/}%
\providecommand \selectlanguage [0]{\@gobble}%
\providecommand \bibinfo  [0]{\@secondoftwo}%
\providecommand \bibfield  [0]{\@secondoftwo}%
\providecommand \translation [1]{[#1]}%
\providecommand \BibitemOpen [0]{}%
\providecommand \bibitemStop [0]{}%
\providecommand \bibitemNoStop [0]{.\EOS\space}%
\providecommand \EOS [0]{\spacefactor3000\relax}%
\providecommand \BibitemShut  [1]{\csname bibitem#1\endcsname}%
\let\auto@bib@innerbib\@empty
\bibitem [{\citenamefont {Horodecki}\ \emph {et~al.}(2009)\citenamefont {Horodecki}, \citenamefont {Horodecki}, \citenamefont {Horodecki},\ and\ \citenamefont {Horodecki}}]{Horodecki_review}%
  \BibitemOpen
  \bibfield  {author} {\bibinfo {author} {\bibfnamefont {R.}~\bibnamefont {Horodecki}}, \bibinfo {author} {\bibfnamefont {P.}~\bibnamefont {Horodecki}}, \bibinfo {author} {\bibfnamefont {M.}~\bibnamefont {Horodecki}},\ and\ \bibinfo {author} {\bibfnamefont {K.}~\bibnamefont {Horodecki}},\ }\href {https://doi.org/10.1103/RevModPhys.81.865} {\bibfield  {journal} {\bibinfo  {journal} {Rev. Mod. Phys.}\ }\textbf {\bibinfo {volume} {81}},\ \bibinfo {pages} {865} (\bibinfo {year} {2009})}\BibitemShut {NoStop}%
\bibitem [{\citenamefont {G\"{u}hne}\ and\ \citenamefont {Tóth}(2009)}]{Guhne_2009}%
  \BibitemOpen
  \bibfield  {author} {\bibinfo {author} {\bibfnamefont {O.}~\bibnamefont {G\"{u}hne}}\ and\ \bibinfo {author} {\bibfnamefont {G.}~\bibnamefont {Tóth}},\ }\href {https://doi.org/10.1016/j.physrep.2009.02.004} {\bibfield  {journal} {\bibinfo  {journal} {Physics Reports}\ }\textbf {\bibinfo {volume} {474}},\ \bibinfo {pages} {1–75} (\bibinfo {year} {2009})}\BibitemShut {NoStop}%
\bibitem [{\citenamefont {Amico}\ \emph {et~al.}(2008)\citenamefont {Amico}, \citenamefont {Fazio}, \citenamefont {Osterloh},\ and\ \citenamefont {Vedral}}]{Amico_review}%
  \BibitemOpen
  \bibfield  {author} {\bibinfo {author} {\bibfnamefont {L.}~\bibnamefont {Amico}}, \bibinfo {author} {\bibfnamefont {R.}~\bibnamefont {Fazio}}, \bibinfo {author} {\bibfnamefont {A.}~\bibnamefont {Osterloh}},\ and\ \bibinfo {author} {\bibfnamefont {V.}~\bibnamefont {Vedral}},\ }\href {https://doi.org/10.1103/RevModPhys.80.517} {\bibfield  {journal} {\bibinfo  {journal} {Rev. Mod. Phys.}\ }\textbf {\bibinfo {volume} {80}},\ \bibinfo {pages} {517} (\bibinfo {year} {2008})}\BibitemShut {NoStop}%
\bibitem [{\citenamefont {Brunner}\ \emph {et~al.}(2014)\citenamefont {Brunner}, \citenamefont {Cavalcanti}, \citenamefont {Pironio}, \citenamefont {Scarani},\ and\ \citenamefont {Wehner}}]{Brunner_review}%
  \BibitemOpen
  \bibfield  {author} {\bibinfo {author} {\bibfnamefont {N.}~\bibnamefont {Brunner}}, \bibinfo {author} {\bibfnamefont {D.}~\bibnamefont {Cavalcanti}}, \bibinfo {author} {\bibfnamefont {S.}~\bibnamefont {Pironio}}, \bibinfo {author} {\bibfnamefont {V.}~\bibnamefont {Scarani}},\ and\ \bibinfo {author} {\bibfnamefont {S.}~\bibnamefont {Wehner}},\ }\href {https://doi.org/10.1103/RevModPhys.86.419} {\bibfield  {journal} {\bibinfo  {journal} {Rev. Mod. Phys.}\ }\textbf {\bibinfo {volume} {86}},\ \bibinfo {pages} {419} (\bibinfo {year} {2014})}\BibitemShut {NoStop}%
\bibitem [{\citenamefont {Ekert}(1991)}]{Ekert_1991}%
  \BibitemOpen
  \bibfield  {author} {\bibinfo {author} {\bibfnamefont {A.~K.}\ \bibnamefont {Ekert}},\ }\href {https://doi.org/10.1103/PhysRevLett.67.661} {\bibfield  {journal} {\bibinfo  {journal} {Phys. Rev. Lett.}\ }\textbf {\bibinfo {volume} {67}},\ \bibinfo {pages} {661} (\bibinfo {year} {1991})}\BibitemShut {NoStop}%
\bibitem [{\citenamefont {Mayers}\ and\ \citenamefont {Yao}(1998)}]{Mayers_1998}%
  \BibitemOpen
  \bibfield  {author} {\bibinfo {author} {\bibfnamefont {D.}~\bibnamefont {Mayers}}\ and\ \bibinfo {author} {\bibfnamefont {A.}~\bibnamefont {Yao}},\ }in\ \href {https://doi.org/10.1109/sfcs.1998.743501} {\emph {\bibinfo {booktitle} {Proceedings 39th Annual Symposium on Foundations of Computer Science (Cat. No.98CB36280)}}},\ \bibinfo {series and number} {SFCS-98}\ (\bibinfo  {publisher} {IEEE Comput. Soc},\ \bibinfo {year} {1998})\ p.\ \bibinfo {pages} {503–509}\BibitemShut {NoStop}%
\bibitem [{\citenamefont {Barrett}\ \emph {et~al.}(2005)\citenamefont {Barrett}, \citenamefont {Hardy},\ and\ \citenamefont {Kent}}]{Barrett_2005}%
  \BibitemOpen
  \bibfield  {author} {\bibinfo {author} {\bibfnamefont {J.}~\bibnamefont {Barrett}}, \bibinfo {author} {\bibfnamefont {L.}~\bibnamefont {Hardy}},\ and\ \bibinfo {author} {\bibfnamefont {A.}~\bibnamefont {Kent}},\ }\href {https://doi.org/10.1103/PhysRevLett.95.010503} {\bibfield  {journal} {\bibinfo  {journal} {Phys. Rev. Lett.}\ }\textbf {\bibinfo {volume} {95}},\ \bibinfo {pages} {010503} (\bibinfo {year} {2005})}\BibitemShut {NoStop}%
\bibitem [{\citenamefont {Ac\'{\i}n}\ \emph {et~al.}(2007)\citenamefont {Ac\'{\i}n}, \citenamefont {Brunner}, \citenamefont {Gisin}, \citenamefont {Massar}, \citenamefont {Pironio},\ and\ \citenamefont {Scarani}}]{Acin_2007}%
  \BibitemOpen
  \bibfield  {author} {\bibinfo {author} {\bibfnamefont {A.}~\bibnamefont {Ac\'{\i}n}}, \bibinfo {author} {\bibfnamefont {N.}~\bibnamefont {Brunner}}, \bibinfo {author} {\bibfnamefont {N.}~\bibnamefont {Gisin}}, \bibinfo {author} {\bibfnamefont {S.}~\bibnamefont {Massar}}, \bibinfo {author} {\bibfnamefont {S.}~\bibnamefont {Pironio}},\ and\ \bibinfo {author} {\bibfnamefont {V.}~\bibnamefont {Scarani}},\ }\href {https://doi.org/10.1103/PhysRevLett.98.230501} {\bibfield  {journal} {\bibinfo  {journal} {Phys. Rev. Lett.}\ }\textbf {\bibinfo {volume} {98}},\ \bibinfo {pages} {230501} (\bibinfo {year} {2007})}\BibitemShut {NoStop}%
\bibitem [{\citenamefont {Pironio}\ \emph {et~al.}(2009)\citenamefont {Pironio}, \citenamefont {Acín}, \citenamefont {Brunner}, \citenamefont {Gisin}, \citenamefont {Massar},\ and\ \citenamefont {Scarani}}]{Pironio_2009}%
  \BibitemOpen
  \bibfield  {author} {\bibinfo {author} {\bibfnamefont {S.}~\bibnamefont {Pironio}}, \bibinfo {author} {\bibfnamefont {A.}~\bibnamefont {Acín}}, \bibinfo {author} {\bibfnamefont {N.}~\bibnamefont {Brunner}}, \bibinfo {author} {\bibfnamefont {N.}~\bibnamefont {Gisin}}, \bibinfo {author} {\bibfnamefont {S.}~\bibnamefont {Massar}},\ and\ \bibinfo {author} {\bibfnamefont {V.}~\bibnamefont {Scarani}},\ }\href {https://doi.org/10.1088/1367-2630/11/4/045021} {\bibfield  {journal} {\bibinfo  {journal} {New Journal of Physics}\ }\textbf {\bibinfo {volume} {11}},\ \bibinfo {pages} {045021} (\bibinfo {year} {2009})}\BibitemShut {NoStop}%
\bibitem [{\citenamefont {Linden}\ and\ \citenamefont {Popescu}(2001)}]{Linden_2001}%
  \BibitemOpen
  \bibfield  {author} {\bibinfo {author} {\bibfnamefont {N.}~\bibnamefont {Linden}}\ and\ \bibinfo {author} {\bibfnamefont {S.}~\bibnamefont {Popescu}},\ }\href {https://doi.org/10.1103/PhysRevLett.87.047901} {\bibfield  {journal} {\bibinfo  {journal} {Phys. Rev. Lett.}\ }\textbf {\bibinfo {volume} {87}},\ \bibinfo {pages} {047901} (\bibinfo {year} {2001})}\BibitemShut {NoStop}%
\bibitem [{\citenamefont {Jozsa}\ and\ \citenamefont {Linden}(2003)}]{Jozsa_2003}%
  \BibitemOpen
  \bibfield  {author} {\bibinfo {author} {\bibfnamefont {R.}~\bibnamefont {Jozsa}}\ and\ \bibinfo {author} {\bibfnamefont {N.}~\bibnamefont {Linden}},\ }\href {https://doi.org/10.1098/rspa.2002.1097} {\bibfield  {journal} {\bibinfo  {journal} {Proceedings of the Royal Society of London. Series A: Mathematical, Physical and Engineering Sciences}\ }\textbf {\bibinfo {volume} {459}},\ \bibinfo {pages} {2011–2032} (\bibinfo {year} {2003})}\BibitemShut {NoStop}%
\bibitem [{\citenamefont {Harrow}\ and\ \citenamefont {Montanaro}(2017)}]{Harrow_2017}%
  \BibitemOpen
  \bibfield  {author} {\bibinfo {author} {\bibfnamefont {A.~W.}\ \bibnamefont {Harrow}}\ and\ \bibinfo {author} {\bibfnamefont {A.}~\bibnamefont {Montanaro}},\ }\href {https://doi.org/10.1038/nature23458} {\bibfield  {journal} {\bibinfo  {journal} {Nature}\ }\textbf {\bibinfo {volume} {549}},\ \bibinfo {pages} {203–209} (\bibinfo {year} {2017})}\BibitemShut {NoStop}%
\bibitem [{\citenamefont {Bauer}\ \emph {et~al.}(2020)\citenamefont {Bauer}, \citenamefont {Bravyi}, \citenamefont {Motta},\ and\ \citenamefont {Chan}}]{Bauer_2020}%
  \BibitemOpen
  \bibfield  {author} {\bibinfo {author} {\bibfnamefont {B.}~\bibnamefont {Bauer}}, \bibinfo {author} {\bibfnamefont {S.}~\bibnamefont {Bravyi}}, \bibinfo {author} {\bibfnamefont {M.}~\bibnamefont {Motta}},\ and\ \bibinfo {author} {\bibfnamefont {G.~K.-L.}\ \bibnamefont {Chan}},\ }\href {https://doi.org/10.1021/acs.chemrev.9b00829} {\bibfield  {journal} {\bibinfo  {journal} {Chemical Reviews}\ }\textbf {\bibinfo {volume} {120}},\ \bibinfo {pages} {12685–12717} (\bibinfo {year} {2020})}\BibitemShut {NoStop}%
\bibitem [{\citenamefont {Giovannetti}\ \emph {et~al.}(2004)\citenamefont {Giovannetti}, \citenamefont {Lloyd},\ and\ \citenamefont {Maccone}}]{Giovannetti_2004}%
  \BibitemOpen
  \bibfield  {author} {\bibinfo {author} {\bibfnamefont {V.}~\bibnamefont {Giovannetti}}, \bibinfo {author} {\bibfnamefont {S.}~\bibnamefont {Lloyd}},\ and\ \bibinfo {author} {\bibfnamefont {L.}~\bibnamefont {Maccone}},\ }\href {https://doi.org/10.1126/science.1104149} {\bibfield  {journal} {\bibinfo  {journal} {Science}\ }\textbf {\bibinfo {volume} {306}},\ \bibinfo {pages} {1330} (\bibinfo {year} {2004})},\ \Eprint {https://arxiv.org/abs/https://www.science.org/doi/pdf/10.1126/science.1104149} {https://www.science.org/doi/pdf/10.1126/science.1104149} \BibitemShut {NoStop}%
\bibitem [{\citenamefont {Giovannetti}\ \emph {et~al.}(2011)\citenamefont {Giovannetti}, \citenamefont {Lloyd},\ and\ \citenamefont {Maccone}}]{Giovannetti_2011}%
  \BibitemOpen
  \bibfield  {author} {\bibinfo {author} {\bibfnamefont {V.}~\bibnamefont {Giovannetti}}, \bibinfo {author} {\bibfnamefont {S.}~\bibnamefont {Lloyd}},\ and\ \bibinfo {author} {\bibfnamefont {L.}~\bibnamefont {Maccone}},\ }\href {https://doi.org/10.1038/nphoton.2011.35} {\bibfield  {journal} {\bibinfo  {journal} {Nature Photonics}\ }\textbf {\bibinfo {volume} {5}},\ \bibinfo {pages} {222–229} (\bibinfo {year} {2011})}\BibitemShut {NoStop}%
\bibitem [{\citenamefont {Wiseman}\ and\ \citenamefont {Milburn}(2009)}]{Wiseman_2009}%
  \BibitemOpen
  \bibfield  {author} {\bibinfo {author} {\bibfnamefont {H.~M.}\ \bibnamefont {Wiseman}}\ and\ \bibinfo {author} {\bibfnamefont {G.~J.}\ \bibnamefont {Milburn}},\ }\href {https://doi.org/10.1017/cbo9780511813948} {\emph {\bibinfo {title} {Quantum Measurement and Control}}}\ (\bibinfo  {publisher} {Cambridge University Press},\ \bibinfo {year} {2009})\BibitemShut {NoStop}%
\bibitem [{\citenamefont {Tóth}\ and\ \citenamefont {Apellaniz}(2014)}]{Toth_2014}%
  \BibitemOpen
  \bibfield  {author} {\bibinfo {author} {\bibfnamefont {G.}~\bibnamefont {Tóth}}\ and\ \bibinfo {author} {\bibfnamefont {I.}~\bibnamefont {Apellaniz}},\ }\href {https://doi.org/10.1088/1751-8113/47/42/424006} {\bibfield  {journal} {\bibinfo  {journal} {Journal of Physics A: Mathematical and Theoretical}\ }\textbf {\bibinfo {volume} {47}},\ \bibinfo {pages} {424006} (\bibinfo {year} {2014})}\BibitemShut {NoStop}%
\bibitem [{\citenamefont {Montenegro}\ \emph {et~al.}(2025)\citenamefont {Montenegro}, \citenamefont {Mukhopadhyay}, \citenamefont {Yousefjani}, \citenamefont {Sarkar}, \citenamefont {Mishra}, \citenamefont {Paris},\ and\ \citenamefont {Bayat}}]{Montenegro_2025}%
  \BibitemOpen
  \bibfield  {author} {\bibinfo {author} {\bibfnamefont {V.}~\bibnamefont {Montenegro}}, \bibinfo {author} {\bibfnamefont {C.}~\bibnamefont {Mukhopadhyay}}, \bibinfo {author} {\bibfnamefont {R.}~\bibnamefont {Yousefjani}}, \bibinfo {author} {\bibfnamefont {S.}~\bibnamefont {Sarkar}}, \bibinfo {author} {\bibfnamefont {U.}~\bibnamefont {Mishra}}, \bibinfo {author} {\bibfnamefont {M.~G.}\ \bibnamefont {Paris}},\ and\ \bibinfo {author} {\bibfnamefont {A.}~\bibnamefont {Bayat}},\ }\href {https://doi.org/10.1016/j.physrep.2025.05.005} {\bibfield  {journal} {\bibinfo  {journal} {Physics Reports}\ }\textbf {\bibinfo {volume} {1134}},\ \bibinfo {pages} {1–62} (\bibinfo {year} {2025})}\BibitemShut {NoStop}%
\bibitem [{\citenamefont {Gottesman}(1997)}]{Gottesman_1997}%
  \BibitemOpen
  \bibfield  {author} {\bibinfo {author} {\bibfnamefont {D.}~\bibnamefont {Gottesman}},\ }\href {https://doi.org/10.48550/ARXIV.QUANT-PH/9705052} {\bibinfo {title} {Stabilizer codes and quantum error correction}} (\bibinfo {year} {1997})\BibitemShut {NoStop}%
\bibitem [{\citenamefont {Gottesman}(1998)}]{Gottesman_1998}%
  \BibitemOpen
  \bibfield  {author} {\bibinfo {author} {\bibfnamefont {D.}~\bibnamefont {Gottesman}},\ }\href {https://arxiv.org/abs/quant-ph/9807006} {\bibinfo {title} {The heisenberg representation of quantum computers}} (\bibinfo {year} {1998}),\ \Eprint {https://arxiv.org/abs/quant-ph/9807006} {arXiv:quant-ph/9807006 [quant-ph]} \BibitemShut {NoStop}%
\bibitem [{\citenamefont {Aaronson}\ and\ \citenamefont {Gottesman}(2004)}]{Aaronson_2004}%
  \BibitemOpen
  \bibfield  {author} {\bibinfo {author} {\bibfnamefont {S.}~\bibnamefont {Aaronson}}\ and\ \bibinfo {author} {\bibfnamefont {D.}~\bibnamefont {Gottesman}},\ }\href {https://doi.org/10.1103/PhysRevA.70.052328} {\bibfield  {journal} {\bibinfo  {journal} {Phys. Rev. A}\ }\textbf {\bibinfo {volume} {70}},\ \bibinfo {pages} {052328} (\bibinfo {year} {2004})}\BibitemShut {NoStop}%
\bibitem [{\citenamefont {Dowling}\ \emph {et~al.}(2004)\citenamefont {Dowling}, \citenamefont {Doherty},\ and\ \citenamefont {Bartlett}}]{Dowling_2004}%
  \BibitemOpen
  \bibfield  {author} {\bibinfo {author} {\bibfnamefont {M.~R.}\ \bibnamefont {Dowling}}, \bibinfo {author} {\bibfnamefont {A.~C.}\ \bibnamefont {Doherty}},\ and\ \bibinfo {author} {\bibfnamefont {S.~D.}\ \bibnamefont {Bartlett}},\ }\href {https://doi.org/10.1103/PhysRevA.70.062113} {\bibfield  {journal} {\bibinfo  {journal} {Phys. Rev. A}\ }\textbf {\bibinfo {volume} {70}},\ \bibinfo {pages} {062113} (\bibinfo {year} {2004})}\BibitemShut {NoStop}%
\bibitem [{\citenamefont {G\"{u}hne}\ \emph {et~al.}(2005)\citenamefont {G\"{u}hne}, \citenamefont {Tóth},\ and\ \citenamefont {Briegel}}]{Guhne_2005}%
  \BibitemOpen
  \bibfield  {author} {\bibinfo {author} {\bibfnamefont {O.}~\bibnamefont {G\"{u}hne}}, \bibinfo {author} {\bibfnamefont {G.}~\bibnamefont {Tóth}},\ and\ \bibinfo {author} {\bibfnamefont {H.~J.}\ \bibnamefont {Briegel}},\ }\href {https://doi.org/10.1088/1367-2630/7/1/229} {\bibfield  {journal} {\bibinfo  {journal} {New Journal of Physics}\ }\textbf {\bibinfo {volume} {7}},\ \bibinfo {pages} {229–229} (\bibinfo {year} {2005})}\BibitemShut {NoStop}%
\bibitem [{\citenamefont {Hofmann}\ \emph {et~al.}(2014)\citenamefont {Hofmann}, \citenamefont {Osterloh},\ and\ \citenamefont {G\"uhne}}]{Hofmann_2014}%
  \BibitemOpen
  \bibfield  {author} {\bibinfo {author} {\bibfnamefont {M.}~\bibnamefont {Hofmann}}, \bibinfo {author} {\bibfnamefont {A.}~\bibnamefont {Osterloh}},\ and\ \bibinfo {author} {\bibfnamefont {O.}~\bibnamefont {G\"uhne}},\ }\href {https://doi.org/10.1103/PhysRevB.89.134101} {\bibfield  {journal} {\bibinfo  {journal} {Phys. Rev. B}\ }\textbf {\bibinfo {volume} {89}},\ \bibinfo {pages} {134101} (\bibinfo {year} {2014})}\BibitemShut {NoStop}%
\bibitem [{\citenamefont {Soldati}\ \emph {et~al.}(2021)\citenamefont {Soldati}, \citenamefont {Mitchison},\ and\ \citenamefont {Landi}}]{Soldati_2021}%
  \BibitemOpen
  \bibfield  {author} {\bibinfo {author} {\bibfnamefont {R.~R.}\ \bibnamefont {Soldati}}, \bibinfo {author} {\bibfnamefont {M.~T.}\ \bibnamefont {Mitchison}},\ and\ \bibinfo {author} {\bibfnamefont {G.~T.}\ \bibnamefont {Landi}},\ }\href {https://doi.org/10.1103/PhysRevA.104.052423} {\bibfield  {journal} {\bibinfo  {journal} {Phys. Rev. A}\ }\textbf {\bibinfo {volume} {104}},\ \bibinfo {pages} {052423} (\bibinfo {year} {2021})}\BibitemShut {NoStop}%
\bibitem [{\citenamefont {Su}\ \emph {et~al.}(2022)\citenamefont {Su}, \citenamefont {Ren}, \citenamefont {Wang},\ and\ \citenamefont {Bai}}]{Su_2022}%
  \BibitemOpen
  \bibfield  {author} {\bibinfo {author} {\bibfnamefont {L.-L.}\ \bibnamefont {Su}}, \bibinfo {author} {\bibfnamefont {J.}~\bibnamefont {Ren}}, \bibinfo {author} {\bibfnamefont {Z.~D.}\ \bibnamefont {Wang}},\ and\ \bibinfo {author} {\bibfnamefont {Y.-K.}\ \bibnamefont {Bai}},\ }\href {https://doi.org/10.1103/PhysRevA.106.042427} {\bibfield  {journal} {\bibinfo  {journal} {Phys. Rev. A}\ }\textbf {\bibinfo {volume} {106}},\ \bibinfo {pages} {042427} (\bibinfo {year} {2022})}\BibitemShut {NoStop}%
\bibitem [{\citenamefont {Cieśliński}\ \emph {et~al.}(2023)\citenamefont {Cieśliński}, \citenamefont {Kłobus}, \citenamefont {Kurzyński}, \citenamefont {Paterek},\ and\ \citenamefont {Laskowski}}]{Cieslinski_2023}%
  \BibitemOpen
  \bibfield  {author} {\bibinfo {author} {\bibfnamefont {P.}~\bibnamefont {Cieśliński}}, \bibinfo {author} {\bibfnamefont {W.}~\bibnamefont {Kłobus}}, \bibinfo {author} {\bibfnamefont {P.}~\bibnamefont {Kurzyński}}, \bibinfo {author} {\bibfnamefont {T.}~\bibnamefont {Paterek}},\ and\ \bibinfo {author} {\bibfnamefont {W.}~\bibnamefont {Laskowski}},\ }\href {https://doi.org/10.1088/1367-2630/acf953} {\bibfield  {journal} {\bibinfo  {journal} {New Journal of Physics}\ }\textbf {\bibinfo {volume} {25}},\ \bibinfo {pages} {093040} (\bibinfo {year} {2023})}\BibitemShut {NoStop}%
\bibitem [{\citenamefont {Cie\ifmmode \acute{s}\else \'{s}\fi{}li\ifmmode~\acute{n}\else \'{n}\fi{}ski}\ \emph {et~al.}(2024)\citenamefont {Cie\ifmmode \acute{s}\else \'{s}\fi{}li\ifmmode~\acute{n}\else \'{n}\fi{}ski}, \citenamefont {Kurzy\ifmmode~\acute{n}\else \'{n}\fi{}ski}, \citenamefont {Sowi\ifmmode~\acute{n}\else \'{n}\fi{}ski}, \citenamefont {K\l{}obus},\ and\ \citenamefont {Laskowski}}]{Cieslinski_2024_fisher}%
  \BibitemOpen
  \bibfield  {author} {\bibinfo {author} {\bibfnamefont {P.}~\bibnamefont {Cie\ifmmode \acute{s}\else \'{s}\fi{}li\ifmmode~\acute{n}\else \'{n}\fi{}ski}}, \bibinfo {author} {\bibfnamefont {P.}~\bibnamefont {Kurzy\ifmmode~\acute{n}\else \'{n}\fi{}ski}}, \bibinfo {author} {\bibfnamefont {T.}~\bibnamefont {Sowi\ifmmode~\acute{n}\else \'{n}\fi{}ski}}, \bibinfo {author} {\bibfnamefont {W.}~\bibnamefont {K\l{}obus}},\ and\ \bibinfo {author} {\bibfnamefont {W.}~\bibnamefont {Laskowski}},\ }\href {https://doi.org/10.1103/PhysRevA.110.012407} {\bibfield  {journal} {\bibinfo  {journal} {Phys. Rev. A}\ }\textbf {\bibinfo {volume} {110}},\ \bibinfo {pages} {012407} (\bibinfo {year} {2024})}\BibitemShut {NoStop}%
\bibitem [{\citenamefont {Consiglio}\ \emph {et~al.}(2025)\citenamefont {Consiglio}, \citenamefont {Odavi\ifmmode~\acute{c}\else \'{c}\fi{}}, \citenamefont {Bonsignori}, \citenamefont {Torre}, \citenamefont {Wie\ifmmode~\acute{s}\else \'{s}\fi{}niak}, \citenamefont {Franchini}, \citenamefont {Giampaolo},\ and\ \citenamefont {Apollaro}}]{Consiglio_2025}%
  \BibitemOpen
  \bibfield  {author} {\bibinfo {author} {\bibfnamefont {M.}~\bibnamefont {Consiglio}}, \bibinfo {author} {\bibfnamefont {J.}~\bibnamefont {Odavi\ifmmode~\acute{c}\else \'{c}\fi{}}}, \bibinfo {author} {\bibfnamefont {R.}~\bibnamefont {Bonsignori}}, \bibinfo {author} {\bibfnamefont {G.}~\bibnamefont {Torre}}, \bibinfo {author} {\bibfnamefont {M.}~\bibnamefont {Wie\ifmmode~\acute{s}\else \'{s}\fi{}niak}}, \bibinfo {author} {\bibfnamefont {F.}~\bibnamefont {Franchini}}, \bibinfo {author} {\bibfnamefont {S.~M.}\ \bibnamefont {Giampaolo}},\ and\ \bibinfo {author} {\bibfnamefont {T.~J.~G.}\ \bibnamefont {Apollaro}},\ }\href {https://doi.org/10.1103/PhysRevA.111.032434} {\bibfield  {journal} {\bibinfo  {journal} {Phys. Rev. A}\ }\textbf {\bibinfo {volume} {111}},\ \bibinfo {pages} {032434} (\bibinfo {year} {2025})}\BibitemShut {NoStop}%
\bibitem [{\citenamefont {Wie\ifmmode~\acute{s}\else \'{s}\fi{}niak}\ \emph {et~al.}(2025)\citenamefont {Wie\ifmmode~\acute{s}\else \'{s}\fi{}niak}, \citenamefont {Kumar},\ and\ \citenamefont {Nkouatchoua~Ngueya}}]{Wiesniak_2025}%
  \BibitemOpen
  \bibfield  {author} {\bibinfo {author} {\bibfnamefont {M.}~\bibnamefont {Wie\ifmmode~\acute{s}\else \'{s}\fi{}niak}}, \bibinfo {author} {\bibfnamefont {A.}~\bibnamefont {Kumar}},\ and\ \bibinfo {author} {\bibfnamefont {I.~H.}\ \bibnamefont {Nkouatchoua~Ngueya}},\ }\href {https://doi.org/10.1103/wwr6-l3dq} {\bibfield  {journal} {\bibinfo  {journal} {Phys. Rev. B}\ }\textbf {\bibinfo {volume} {112}},\ \bibinfo {pages} {134425} (\bibinfo {year} {2025})}\BibitemShut {NoStop}%
\bibitem [{\citenamefont {Markham}\ \emph {et~al.}(2008)\citenamefont {Markham}, \citenamefont {Anders}, \citenamefont {Vedral}, \citenamefont {Murao},\ and\ \citenamefont {Miyake}}]{Markham_2008}%
  \BibitemOpen
  \bibfield  {author} {\bibinfo {author} {\bibfnamefont {D.}~\bibnamefont {Markham}}, \bibinfo {author} {\bibfnamefont {J.}~\bibnamefont {Anders}}, \bibinfo {author} {\bibfnamefont {V.}~\bibnamefont {Vedral}}, \bibinfo {author} {\bibfnamefont {M.}~\bibnamefont {Murao}},\ and\ \bibinfo {author} {\bibfnamefont {A.}~\bibnamefont {Miyake}},\ }\href {https://doi.org/10.1209/0295-5075/81/40006} {\bibfield  {journal} {\bibinfo  {journal} {EPL (Europhysics Letters)}\ }\textbf {\bibinfo {volume} {81}},\ \bibinfo {pages} {40006} (\bibinfo {year} {2008})}\BibitemShut {NoStop}%
\bibitem [{\citenamefont {Nakata}\ \emph {et~al.}(2009)\citenamefont {Nakata}, \citenamefont {Markham},\ and\ \citenamefont {Murao}}]{Nakata_2009}%
  \BibitemOpen
  \bibfield  {author} {\bibinfo {author} {\bibfnamefont {Y.}~\bibnamefont {Nakata}}, \bibinfo {author} {\bibfnamefont {D.}~\bibnamefont {Markham}},\ and\ \bibinfo {author} {\bibfnamefont {M.}~\bibnamefont {Murao}},\ }\href {https://doi.org/10.1103/PhysRevA.79.042313} {\bibfield  {journal} {\bibinfo  {journal} {Phys. Rev. A}\ }\textbf {\bibinfo {volume} {79}},\ \bibinfo {pages} {042313} (\bibinfo {year} {2009})}\BibitemShut {NoStop}%
\bibitem [{\citenamefont {Anders}\ and\ \citenamefont {Vedral}(2007)}]{Anders_2007}%
  \BibitemOpen
  \bibfield  {author} {\bibinfo {author} {\bibfnamefont {J.}~\bibnamefont {Anders}}\ and\ \bibinfo {author} {\bibfnamefont {V.}~\bibnamefont {Vedral}},\ }\href {https://doi.org/10.1007/s11080-007-9034-6} {\bibfield  {journal} {\bibinfo  {journal} {Open Systems \&; Information Dynamics}\ }\textbf {\bibinfo {volume} {14}},\ \bibinfo {pages} {1–16} (\bibinfo {year} {2007})}\BibitemShut {NoStop}%
\bibitem [{\citenamefont {Li}\ \emph {et~al.}(2024)\citenamefont {Li}, \citenamefont {Zhou}, \citenamefont {Zhang}, \citenamefont {Bai},\ and\ \citenamefont {Lin}}]{Li_2024}%
  \BibitemOpen
  \bibfield  {author} {\bibinfo {author} {\bibfnamefont {Y.-C.}\ \bibnamefont {Li}}, \bibinfo {author} {\bibfnamefont {Y.-H.}\ \bibnamefont {Zhou}}, \bibinfo {author} {\bibfnamefont {Y.}~\bibnamefont {Zhang}}, \bibinfo {author} {\bibfnamefont {Y.-K.}\ \bibnamefont {Bai}},\ and\ \bibinfo {author} {\bibfnamefont {H.-Q.}\ \bibnamefont {Lin}},\ }\href {https://doi.org/10.1088/1367-2630/ad273a} {\bibfield  {journal} {\bibinfo  {journal} {New Journal of Physics}\ }\textbf {\bibinfo {volume} {26}},\ \bibinfo {pages} {023031} (\bibinfo {year} {2024})}\BibitemShut {NoStop}%
\bibitem [{\citenamefont {Liu}\ and\ \citenamefont {Winter}(2022)}]{Liu_2022}%
  \BibitemOpen
  \bibfield  {author} {\bibinfo {author} {\bibfnamefont {Z.-W.}\ \bibnamefont {Liu}}\ and\ \bibinfo {author} {\bibfnamefont {A.}~\bibnamefont {Winter}},\ }\href {https://doi.org/10.1103/PRXQuantum.3.020333} {\bibfield  {journal} {\bibinfo  {journal} {PRX Quantum}\ }\textbf {\bibinfo {volume} {3}},\ \bibinfo {pages} {020333} (\bibinfo {year} {2022})}\BibitemShut {NoStop}%
\bibitem [{\citenamefont {Fu}\ \emph {et~al.}(2022)\citenamefont {Fu}, \citenamefont {Li},\ and\ \citenamefont {Luo}}]{Fu_2022}%
  \BibitemOpen
  \bibfield  {author} {\bibinfo {author} {\bibfnamefont {S.}~\bibnamefont {Fu}}, \bibinfo {author} {\bibfnamefont {X.}~\bibnamefont {Li}},\ and\ \bibinfo {author} {\bibfnamefont {S.}~\bibnamefont {Luo}},\ }\href {https://doi.org/10.1103/PhysRevA.106.062405} {\bibfield  {journal} {\bibinfo  {journal} {Phys. Rev. A}\ }\textbf {\bibinfo {volume} {106}},\ \bibinfo {pages} {062405} (\bibinfo {year} {2022})}\BibitemShut {NoStop}%
\bibitem [{\citenamefont {Haug}\ and\ \citenamefont {Piroli}(2023)}]{Huang_2023}%
  \BibitemOpen
  \bibfield  {author} {\bibinfo {author} {\bibfnamefont {T.}~\bibnamefont {Haug}}\ and\ \bibinfo {author} {\bibfnamefont {L.}~\bibnamefont {Piroli}},\ }\href {https://doi.org/10.1103/PhysRevB.107.035148} {\bibfield  {journal} {\bibinfo  {journal} {Phys. Rev. B}\ }\textbf {\bibinfo {volume} {107}},\ \bibinfo {pages} {035148} (\bibinfo {year} {2023})}\BibitemShut {NoStop}%
\bibitem [{\citenamefont {Tarabunga}(2024)}]{Tarabunga_2024}%
  \BibitemOpen
  \bibfield  {author} {\bibinfo {author} {\bibfnamefont {P.~S.}\ \bibnamefont {Tarabunga}},\ }\href {https://doi.org/10.22331/q-2024-07-17-1413} {\bibfield  {journal} {\bibinfo  {journal} {Quantum}\ }\textbf {\bibinfo {volume} {8}},\ \bibinfo {pages} {1413} (\bibinfo {year} {2024})}\BibitemShut {NoStop}%
\bibitem [{\citenamefont {Defenu}\ \emph {et~al.}(2023)\citenamefont {Defenu}, \citenamefont {Donner}, \citenamefont {Macr\`{\i}}, \citenamefont {Pagano}, \citenamefont {Ruffo},\ and\ \citenamefont {Trombettoni}}]{longrange_review}%
  \BibitemOpen
  \bibfield  {author} {\bibinfo {author} {\bibfnamefont {N.}~\bibnamefont {Defenu}}, \bibinfo {author} {\bibfnamefont {T.}~\bibnamefont {Donner}}, \bibinfo {author} {\bibfnamefont {T.}~\bibnamefont {Macr\`{\i}}}, \bibinfo {author} {\bibfnamefont {G.}~\bibnamefont {Pagano}}, \bibinfo {author} {\bibfnamefont {S.}~\bibnamefont {Ruffo}},\ and\ \bibinfo {author} {\bibfnamefont {A.}~\bibnamefont {Trombettoni}},\ }\href {https://doi.org/10.1103/RevModPhys.95.035002} {\bibfield  {journal} {\bibinfo  {journal} {Rev. Mod. Phys.}\ }\textbf {\bibinfo {volume} {95}},\ \bibinfo {pages} {035002} (\bibinfo {year} {2023})}\BibitemShut {NoStop}%
\bibitem [{\citenamefont {Lahaye}\ \emph {et~al.}(2009)\citenamefont {Lahaye}, \citenamefont {Menotti}, \citenamefont {Santos}, \citenamefont {Lewenstein},\ and\ \citenamefont {Pfau}}]{Lahaye_2009}%
  \BibitemOpen
  \bibfield  {author} {\bibinfo {author} {\bibfnamefont {T.}~\bibnamefont {Lahaye}}, \bibinfo {author} {\bibfnamefont {C.}~\bibnamefont {Menotti}}, \bibinfo {author} {\bibfnamefont {L.}~\bibnamefont {Santos}}, \bibinfo {author} {\bibfnamefont {M.}~\bibnamefont {Lewenstein}},\ and\ \bibinfo {author} {\bibfnamefont {T.}~\bibnamefont {Pfau}},\ }\href {https://doi.org/10.1088/0034-4885/72/12/126401} {\bibfield  {journal} {\bibinfo  {journal} {Reports on Progress in Physics}\ }\textbf {\bibinfo {volume} {72}},\ \bibinfo {pages} {126401} (\bibinfo {year} {2009})}\BibitemShut {NoStop}%
\bibitem [{\citenamefont {Koffel}\ \emph {et~al.}(2012)\citenamefont {Koffel}, \citenamefont {Lewenstein},\ and\ \citenamefont {Tagliacozzo}}]{Koffel_2012}%
  \BibitemOpen
  \bibfield  {author} {\bibinfo {author} {\bibfnamefont {T.}~\bibnamefont {Koffel}}, \bibinfo {author} {\bibfnamefont {M.}~\bibnamefont {Lewenstein}},\ and\ \bibinfo {author} {\bibfnamefont {L.}~\bibnamefont {Tagliacozzo}},\ }\href {https://doi.org/10.1103/PhysRevLett.109.267203} {\bibfield  {journal} {\bibinfo  {journal} {Phys. Rev. Lett.}\ }\textbf {\bibinfo {volume} {109}},\ \bibinfo {pages} {267203} (\bibinfo {year} {2012})}\BibitemShut {NoStop}%
\bibitem [{\citenamefont {Knap}\ \emph {et~al.}(2013)\citenamefont {Knap}, \citenamefont {Kantian}, \citenamefont {Giamarchi}, \citenamefont {Bloch}, \citenamefont {Lukin},\ and\ \citenamefont {Demler}}]{Knap_2013}%
  \BibitemOpen
  \bibfield  {author} {\bibinfo {author} {\bibfnamefont {M.}~\bibnamefont {Knap}}, \bibinfo {author} {\bibfnamefont {A.}~\bibnamefont {Kantian}}, \bibinfo {author} {\bibfnamefont {T.}~\bibnamefont {Giamarchi}}, \bibinfo {author} {\bibfnamefont {I.}~\bibnamefont {Bloch}}, \bibinfo {author} {\bibfnamefont {M.~D.}\ \bibnamefont {Lukin}},\ and\ \bibinfo {author} {\bibfnamefont {E.}~\bibnamefont {Demler}},\ }\bibfield  {journal} {\bibinfo  {journal} {Physical Review Letters}\ }\textbf {\bibinfo {volume} {111}},\ \href {https://doi.org/10.1103/physrevlett.111.147205} {10.1103/physrevlett.111.147205} (\bibinfo {year} {2013})\BibitemShut {NoStop}%
\bibitem [{\citenamefont {Vodola}\ \emph {et~al.}(2015)\citenamefont {Vodola}, \citenamefont {Lepori}, \citenamefont {Ercolessi},\ and\ \citenamefont {Pupillo}}]{Vodola_2015}%
  \BibitemOpen
  \bibfield  {author} {\bibinfo {author} {\bibfnamefont {D.}~\bibnamefont {Vodola}}, \bibinfo {author} {\bibfnamefont {L.}~\bibnamefont {Lepori}}, \bibinfo {author} {\bibfnamefont {E.}~\bibnamefont {Ercolessi}},\ and\ \bibinfo {author} {\bibfnamefont {G.}~\bibnamefont {Pupillo}},\ }\href {https://doi.org/10.1088/1367-2630/18/1/015001} {\bibfield  {journal} {\bibinfo  {journal} {New Journal of Physics}\ }\textbf {\bibinfo {volume} {18}},\ \bibinfo {pages} {015001} (\bibinfo {year} {2015})}\BibitemShut {NoStop}%
\bibitem [{\citenamefont {Leone}\ \emph {et~al.}(2022)\citenamefont {Leone}, \citenamefont {Oliviero},\ and\ \citenamefont {Hamma}}]{Leone_2022}%
  \BibitemOpen
  \bibfield  {author} {\bibinfo {author} {\bibfnamefont {L.}~\bibnamefont {Leone}}, \bibinfo {author} {\bibfnamefont {S.~F.}\ \bibnamefont {Oliviero}},\ and\ \bibinfo {author} {\bibfnamefont {A.}~\bibnamefont {Hamma}},\ }\bibfield  {journal} {\bibinfo  {journal} {Physical Review Letters}\ }\textbf {\bibinfo {volume} {128}},\ \href {https://doi.org/10.1103/physrevlett.128.050402} {10.1103/physrevlett.128.050402} (\bibinfo {year} {2022})\BibitemShut {NoStop}%
\bibitem [{\citenamefont {Coffman}\ \emph {et~al.}(2000)\citenamefont {Coffman}, \citenamefont {Kundu},\ and\ \citenamefont {Wootters}}]{Coffman_2000}%
  \BibitemOpen
  \bibfield  {author} {\bibinfo {author} {\bibfnamefont {V.}~\bibnamefont {Coffman}}, \bibinfo {author} {\bibfnamefont {J.}~\bibnamefont {Kundu}},\ and\ \bibinfo {author} {\bibfnamefont {W.~K.}\ \bibnamefont {Wootters}},\ }\href {https://doi.org/10.1103/PhysRevA.61.052306} {\bibfield  {journal} {\bibinfo  {journal} {Phys. Rev. A}\ }\textbf {\bibinfo {volume} {61}},\ \bibinfo {pages} {052306} (\bibinfo {year} {2000})}\BibitemShut {NoStop}%
\bibitem [{\citenamefont {Hassan}\ and\ \citenamefont {Joag}(2008)}]{Hassan_2008}%
  \BibitemOpen
  \bibfield  {author} {\bibinfo {author} {\bibfnamefont {A.~S.~M.}\ \bibnamefont {Hassan}}\ and\ \bibinfo {author} {\bibfnamefont {P.~S.}\ \bibnamefont {Joag}},\ }\href {https://doi.org/10.1103/PhysRevA.77.062334} {\bibfield  {journal} {\bibinfo  {journal} {Phys. Rev. A}\ }\textbf {\bibinfo {volume} {77}},\ \bibinfo {pages} {062334} (\bibinfo {year} {2008})}\BibitemShut {NoStop}%
\bibitem [{\citenamefont {Badzia\ifmmode~\mbox{\c{}}\else \c{}\fi{}g}\ \emph {et~al.}(2008)\citenamefont {Badzia\ifmmode~\mbox{\c{}}\else \c{}\fi{}g}, \citenamefont {Brukner}, \citenamefont {Laskowski}, \citenamefont {Paterek},\ and\ \citenamefont {\ifmmode~\dot{Z}\else \.{Z}\fi{}ukowski}}]{Badziag_2008}%
  \BibitemOpen
  \bibfield  {author} {\bibinfo {author} {\bibfnamefont {P.}~\bibnamefont {Badzia\ifmmode~\mbox{\c{}}\else \c{}\fi{}g}}, \bibinfo {author} {\bibfnamefont {i.~c.~v.}\ \bibnamefont {Brukner}}, \bibinfo {author} {\bibfnamefont {W.}~\bibnamefont {Laskowski}}, \bibinfo {author} {\bibfnamefont {T.}~\bibnamefont {Paterek}},\ and\ \bibinfo {author} {\bibfnamefont {M.}~\bibnamefont {\ifmmode~\dot{Z}\else \.{Z}\fi{}ukowski}},\ }\href {https://doi.org/10.1103/PhysRevLett.100.140403} {\bibfield  {journal} {\bibinfo  {journal} {Phys. Rev. Lett.}\ }\textbf {\bibinfo {volume} {100}},\ \bibinfo {pages} {140403} (\bibinfo {year} {2008})}\BibitemShut {NoStop}%
\bibitem [{\citenamefont {Tran}\ \emph {et~al.}(2015)\citenamefont {Tran}, \citenamefont {Daki\ifmmode~\acute{c}\else \'{c}\fi{}}, \citenamefont {Arnault}, \citenamefont {Laskowski},\ and\ \citenamefont {Paterek}}]{Tran_2015}%
  \BibitemOpen
  \bibfield  {author} {\bibinfo {author} {\bibfnamefont {M.~C.}\ \bibnamefont {Tran}}, \bibinfo {author} {\bibfnamefont {B.}~\bibnamefont {Daki\ifmmode~\acute{c}\else \'{c}\fi{}}}, \bibinfo {author} {\bibfnamefont {F.~m.~c.}\ \bibnamefont {Arnault}}, \bibinfo {author} {\bibfnamefont {W.}~\bibnamefont {Laskowski}},\ and\ \bibinfo {author} {\bibfnamefont {T.}~\bibnamefont {Paterek}},\ }\href {https://doi.org/10.1103/PhysRevA.92.050301} {\bibfield  {journal} {\bibinfo  {journal} {Phys. Rev. A}\ }\textbf {\bibinfo {volume} {92}},\ \bibinfo {pages} {050301} (\bibinfo {year} {2015})}\BibitemShut {NoStop}%
\bibitem [{\citenamefont {Ketterer}\ \emph {et~al.}(2019)\citenamefont {Ketterer}, \citenamefont {Wyderka},\ and\ \citenamefont {G\"uhne}}]{Ketterer_2019}%
  \BibitemOpen
  \bibfield  {author} {\bibinfo {author} {\bibfnamefont {A.}~\bibnamefont {Ketterer}}, \bibinfo {author} {\bibfnamefont {N.}~\bibnamefont {Wyderka}},\ and\ \bibinfo {author} {\bibfnamefont {O.}~\bibnamefont {G\"uhne}},\ }\href {https://doi.org/10.1103/PhysRevLett.122.120505} {\bibfield  {journal} {\bibinfo  {journal} {Phys. Rev. Lett.}\ }\textbf {\bibinfo {volume} {122}},\ \bibinfo {pages} {120505} (\bibinfo {year} {2019})}\BibitemShut {NoStop}%
\bibitem [{\citenamefont {Wyderka}\ and\ \citenamefont {G\"{u}hne}(2020)}]{Wyderka_2020}%
  \BibitemOpen
  \bibfield  {author} {\bibinfo {author} {\bibfnamefont {N.}~\bibnamefont {Wyderka}}\ and\ \bibinfo {author} {\bibfnamefont {O.}~\bibnamefont {G\"{u}hne}},\ }\href {https://doi.org/10.1088/1751-8121/ab7f0a} {\bibfield  {journal} {\bibinfo  {journal} {Journal of Physics A: Mathematical and Theoretical}\ }\textbf {\bibinfo {volume} {53}},\ \bibinfo {pages} {345302} (\bibinfo {year} {2020})}\BibitemShut {NoStop}%
\bibitem [{\citenamefont {\ifmmode~\dot{Z}\else \.{Z}\fi{}ukowski}\ and\ \citenamefont {Brukner}(2002)}]{Zukowski_2002}%
  \BibitemOpen
  \bibfield  {author} {\bibinfo {author} {\bibfnamefont {M.}~\bibnamefont {\ifmmode~\dot{Z}\else \.{Z}\fi{}ukowski}}\ and\ \bibinfo {author} {\bibfnamefont {i.~c.~v.}\ \bibnamefont {Brukner}},\ }\href {https://doi.org/10.1103/PhysRevLett.88.210401} {\bibfield  {journal} {\bibinfo  {journal} {Phys. Rev. Lett.}\ }\textbf {\bibinfo {volume} {88}},\ \bibinfo {pages} {210401} (\bibinfo {year} {2002})}\BibitemShut {NoStop}%
\bibitem [{\citenamefont {Cieśliński}\ \emph {et~al.}(2024)\citenamefont {Cieśliński}, \citenamefont {Imai}, \citenamefont {Dziewior}, \citenamefont {G\"{u}hne}, \citenamefont {Knips}, \citenamefont {Laskowski}, \citenamefont {Meinecke}, \citenamefont {Paterek},\ and\ \citenamefont {Vértesi}}]{Cieslinski_2024}%
  \BibitemOpen
  \bibfield  {author} {\bibinfo {author} {\bibfnamefont {P.}~\bibnamefont {Cieśliński}}, \bibinfo {author} {\bibfnamefont {S.}~\bibnamefont {Imai}}, \bibinfo {author} {\bibfnamefont {J.}~\bibnamefont {Dziewior}}, \bibinfo {author} {\bibfnamefont {O.}~\bibnamefont {G\"{u}hne}}, \bibinfo {author} {\bibfnamefont {L.}~\bibnamefont {Knips}}, \bibinfo {author} {\bibfnamefont {W.}~\bibnamefont {Laskowski}}, \bibinfo {author} {\bibfnamefont {J.}~\bibnamefont {Meinecke}}, \bibinfo {author} {\bibfnamefont {T.}~\bibnamefont {Paterek}},\ and\ \bibinfo {author} {\bibfnamefont {T.}~\bibnamefont {Vértesi}},\ }\href {https://doi.org/10.1016/j.physrep.2024.09.009} {\bibfield  {journal} {\bibinfo  {journal} {Physics Reports}\ }\textbf {\bibinfo {volume} {1095}},\ \bibinfo {pages} {1–48} (\bibinfo {year} {2024})}\BibitemShut {NoStop}%
\bibitem [{\citenamefont {Fey}\ and\ \citenamefont {Schmidt}(2016)}]{Fey_2016}%
  \BibitemOpen
  \bibfield  {author} {\bibinfo {author} {\bibfnamefont {S.}~\bibnamefont {Fey}}\ and\ \bibinfo {author} {\bibfnamefont {K.~P.}\ \bibnamefont {Schmidt}},\ }\href {https://doi.org/10.1103/PhysRevB.94.075156} {\bibfield  {journal} {\bibinfo  {journal} {Phys. Rev. B}\ }\textbf {\bibinfo {volume} {94}},\ \bibinfo {pages} {075156} (\bibinfo {year} {2016})}\BibitemShut {NoStop}%
\bibitem [{\citenamefont {Britton}\ \emph {et~al.}(2012)\citenamefont {Britton}, \citenamefont {Sawyer}, \citenamefont {Keith}, \citenamefont {Wang}, \citenamefont {Freericks}, \citenamefont {Uys}, \citenamefont {Biercuk},\ and\ \citenamefont {Bollinger}}]{Britton_2012}%
  \BibitemOpen
  \bibfield  {author} {\bibinfo {author} {\bibfnamefont {J.~W.}\ \bibnamefont {Britton}}, \bibinfo {author} {\bibfnamefont {B.~C.}\ \bibnamefont {Sawyer}}, \bibinfo {author} {\bibfnamefont {A.~C.}\ \bibnamefont {Keith}}, \bibinfo {author} {\bibfnamefont {C.-C.~J.}\ \bibnamefont {Wang}}, \bibinfo {author} {\bibfnamefont {J.~K.}\ \bibnamefont {Freericks}}, \bibinfo {author} {\bibfnamefont {H.}~\bibnamefont {Uys}}, \bibinfo {author} {\bibfnamefont {M.~J.}\ \bibnamefont {Biercuk}},\ and\ \bibinfo {author} {\bibfnamefont {J.~J.}\ \bibnamefont {Bollinger}},\ }\href {https://doi.org/10.1038/nature10981} {\bibfield  {journal} {\bibinfo  {journal} {Nature}\ }\textbf {\bibinfo {volume} {484}},\ \bibinfo {pages} {489–492} (\bibinfo {year} {2012})}\BibitemShut {NoStop}%
\bibitem [{\citenamefont {Islam}\ \emph {et~al.}(2013)\citenamefont {Islam}, \citenamefont {Senko}, \citenamefont {Campbell}, \citenamefont {Korenblit}, \citenamefont {Smith}, \citenamefont {Lee}, \citenamefont {Edwards}, \citenamefont {Wang}, \citenamefont {Freericks},\ and\ \citenamefont {Monroe}}]{Islam_2013}%
  \BibitemOpen
  \bibfield  {author} {\bibinfo {author} {\bibfnamefont {R.}~\bibnamefont {Islam}}, \bibinfo {author} {\bibfnamefont {C.}~\bibnamefont {Senko}}, \bibinfo {author} {\bibfnamefont {W.~C.}\ \bibnamefont {Campbell}}, \bibinfo {author} {\bibfnamefont {S.}~\bibnamefont {Korenblit}}, \bibinfo {author} {\bibfnamefont {J.}~\bibnamefont {Smith}}, \bibinfo {author} {\bibfnamefont {A.}~\bibnamefont {Lee}}, \bibinfo {author} {\bibfnamefont {E.~E.}\ \bibnamefont {Edwards}}, \bibinfo {author} {\bibfnamefont {C.-C.~J.}\ \bibnamefont {Wang}}, \bibinfo {author} {\bibfnamefont {J.~K.}\ \bibnamefont {Freericks}},\ and\ \bibinfo {author} {\bibfnamefont {C.}~\bibnamefont {Monroe}},\ }\href {https://doi.org/10.1126/science.1232296} {\bibfield  {journal} {\bibinfo  {journal} {Science}\ }\textbf {\bibinfo {volume} {340}},\ \bibinfo {pages} {583–587} (\bibinfo {year} {2013})}\BibitemShut {NoStop}%
\bibitem [{\citenamefont {Bohnet}\ \emph {et~al.}(2016)\citenamefont {Bohnet}, \citenamefont {Sawyer}, \citenamefont {Britton}, \citenamefont {Wall}, \citenamefont {Rey}, \citenamefont {Foss-Feig},\ and\ \citenamefont {Bollinger}}]{Bohnet_2016}%
  \BibitemOpen
  \bibfield  {author} {\bibinfo {author} {\bibfnamefont {J.~G.}\ \bibnamefont {Bohnet}}, \bibinfo {author} {\bibfnamefont {B.~C.}\ \bibnamefont {Sawyer}}, \bibinfo {author} {\bibfnamefont {J.~W.}\ \bibnamefont {Britton}}, \bibinfo {author} {\bibfnamefont {M.~L.}\ \bibnamefont {Wall}}, \bibinfo {author} {\bibfnamefont {A.~M.}\ \bibnamefont {Rey}}, \bibinfo {author} {\bibfnamefont {M.}~\bibnamefont {Foss-Feig}},\ and\ \bibinfo {author} {\bibfnamefont {J.~J.}\ \bibnamefont {Bollinger}},\ }\href {https://doi.org/10.1126/science.aad9958} {\bibfield  {journal} {\bibinfo  {journal} {Science}\ }\textbf {\bibinfo {volume} {352}},\ \bibinfo {pages} {1297–1301} (\bibinfo {year} {2016})}\BibitemShut {NoStop}%
\bibitem [{\citenamefont {Yang}\ \emph {et~al.}(2019)\citenamefont {Yang}, \citenamefont {Jiang},\ and\ \citenamefont {Zhou}}]{Yang_2019}%
  \BibitemOpen
  \bibfield  {author} {\bibinfo {author} {\bibfnamefont {F.}~\bibnamefont {Yang}}, \bibinfo {author} {\bibfnamefont {S.-J.}\ \bibnamefont {Jiang}},\ and\ \bibinfo {author} {\bibfnamefont {F.}~\bibnamefont {Zhou}},\ }\href {https://doi.org/10.1103/PhysRevA.99.012119} {\bibfield  {journal} {\bibinfo  {journal} {Phys. Rev. A}\ }\textbf {\bibinfo {volume} {99}},\ \bibinfo {pages} {012119} (\bibinfo {year} {2019})}\BibitemShut {NoStop}%
\bibitem [{\citenamefont {Hauke}\ \emph {et~al.}(2010)\citenamefont {Hauke}, \citenamefont {Cucchietti}, \citenamefont {M\"{u}ller-Hermes}, \citenamefont {Bañuls}, \citenamefont {Ignacio~Cirac},\ and\ \citenamefont {Lewenstein}}]{Hauke2010}%
  \BibitemOpen
  \bibfield  {author} {\bibinfo {author} {\bibfnamefont {P.}~\bibnamefont {Hauke}}, \bibinfo {author} {\bibfnamefont {F.~M.}\ \bibnamefont {Cucchietti}}, \bibinfo {author} {\bibfnamefont {A.}~\bibnamefont {M\"{u}ller-Hermes}}, \bibinfo {author} {\bibfnamefont {M.-C.}\ \bibnamefont {Bañuls}}, \bibinfo {author} {\bibfnamefont {J.}~\bibnamefont {Ignacio~Cirac}},\ and\ \bibinfo {author} {\bibfnamefont {M.}~\bibnamefont {Lewenstein}},\ }\href {https://doi.org/10.1088/1367-2630/12/11/113037} {\bibfield  {journal} {\bibinfo  {journal} {New Journal of Physics}\ }\textbf {\bibinfo {volume} {12}},\ \bibinfo {pages} {113037} (\bibinfo {year} {2010})}\BibitemShut {NoStop}%
\bibitem [{\citenamefont {Peter}\ \emph {et~al.}(2012)\citenamefont {Peter}, \citenamefont {M\"uller}, \citenamefont {Wessel},\ and\ \citenamefont {B\"uchler}}]{Peter_2012}%
  \BibitemOpen
  \bibfield  {author} {\bibinfo {author} {\bibfnamefont {D.}~\bibnamefont {Peter}}, \bibinfo {author} {\bibfnamefont {S.}~\bibnamefont {M\"uller}}, \bibinfo {author} {\bibfnamefont {S.}~\bibnamefont {Wessel}},\ and\ \bibinfo {author} {\bibfnamefont {H.~P.}\ \bibnamefont {B\"uchler}},\ }\href {https://doi.org/10.1103/PhysRevLett.109.025303} {\bibfield  {journal} {\bibinfo  {journal} {Phys. Rev. Lett.}\ }\textbf {\bibinfo {volume} {109}},\ \bibinfo {pages} {025303} (\bibinfo {year} {2012})}\BibitemShut {NoStop}%
\bibitem [{\citenamefont {Shor}\ and\ \citenamefont {Laflamme}(1997)}]{Shor_1997}%
  \BibitemOpen
  \bibfield  {author} {\bibinfo {author} {\bibfnamefont {P.}~\bibnamefont {Shor}}\ and\ \bibinfo {author} {\bibfnamefont {R.}~\bibnamefont {Laflamme}},\ }\href {https://doi.org/10.1103/PhysRevLett.78.1600} {\bibfield  {journal} {\bibinfo  {journal} {Phys. Rev. Lett.}\ }\textbf {\bibinfo {volume} {78}},\ \bibinfo {pages} {1600} (\bibinfo {year} {1997})}\BibitemShut {NoStop}%
\bibitem [{\citenamefont {Fisher}\ \emph {et~al.}(1972)\citenamefont {Fisher}, \citenamefont {Ma},\ and\ \citenamefont {Nickel}}]{Fisher_1972}%
  \BibitemOpen
  \bibfield  {author} {\bibinfo {author} {\bibfnamefont {M.~E.}\ \bibnamefont {Fisher}}, \bibinfo {author} {\bibfnamefont {S.-k.}\ \bibnamefont {Ma}},\ and\ \bibinfo {author} {\bibfnamefont {B.~G.}\ \bibnamefont {Nickel}},\ }\href {https://doi.org/10.1103/PhysRevLett.29.917} {\bibfield  {journal} {\bibinfo  {journal} {Phys. Rev. Lett.}\ }\textbf {\bibinfo {volume} {29}},\ \bibinfo {pages} {917} (\bibinfo {year} {1972})}\BibitemShut {NoStop}%
\bibitem [{\citenamefont {Dutta}\ and\ \citenamefont {Bhattacharjee}(2001)}]{Dutta_2001}%
  \BibitemOpen
  \bibfield  {author} {\bibinfo {author} {\bibfnamefont {A.}~\bibnamefont {Dutta}}\ and\ \bibinfo {author} {\bibfnamefont {J.~K.}\ \bibnamefont {Bhattacharjee}},\ }\href {https://doi.org/10.1103/PhysRevB.64.184106} {\bibfield  {journal} {\bibinfo  {journal} {Phys. Rev. B}\ }\textbf {\bibinfo {volume} {64}},\ \bibinfo {pages} {184106} (\bibinfo {year} {2001})}\BibitemShut {NoStop}%
\bibitem [{\citenamefont {Defenu}\ \emph {et~al.}(2017)\citenamefont {Defenu}, \citenamefont {Trombettoni},\ and\ \citenamefont {Ruffo}}]{Defenu_2017}%
  \BibitemOpen
  \bibfield  {author} {\bibinfo {author} {\bibfnamefont {N.}~\bibnamefont {Defenu}}, \bibinfo {author} {\bibfnamefont {A.}~\bibnamefont {Trombettoni}},\ and\ \bibinfo {author} {\bibfnamefont {S.}~\bibnamefont {Ruffo}},\ }\href {https://doi.org/10.1103/PhysRevB.96.104432} {\bibfield  {journal} {\bibinfo  {journal} {Phys. Rev. B}\ }\textbf {\bibinfo {volume} {96}},\ \bibinfo {pages} {104432} (\bibinfo {year} {2017})}\BibitemShut {NoStop}%
\bibitem [{\citenamefont {Howard}\ and\ \citenamefont {Vala}(2012)}]{Howard_2012}%
  \BibitemOpen
  \bibfield  {author} {\bibinfo {author} {\bibfnamefont {M.}~\bibnamefont {Howard}}\ and\ \bibinfo {author} {\bibfnamefont {J.}~\bibnamefont {Vala}},\ }\href {https://doi.org/10.1103/PhysRevA.85.022304} {\bibfield  {journal} {\bibinfo  {journal} {Phys. Rev. A}\ }\textbf {\bibinfo {volume} {85}},\ \bibinfo {pages} {022304} (\bibinfo {year} {2012})}\BibitemShut {NoStop}%
\bibitem [{\citenamefont {Howard}\ \emph {et~al.}(2014)\citenamefont {Howard}, \citenamefont {Wallman}, \citenamefont {Veitch},\ and\ \citenamefont {Emerson}}]{Howard_2014}%
  \BibitemOpen
  \bibfield  {author} {\bibinfo {author} {\bibfnamefont {M.}~\bibnamefont {Howard}}, \bibinfo {author} {\bibfnamefont {J.}~\bibnamefont {Wallman}}, \bibinfo {author} {\bibfnamefont {V.}~\bibnamefont {Veitch}},\ and\ \bibinfo {author} {\bibfnamefont {J.}~\bibnamefont {Emerson}},\ }\href {https://doi.org/10.1038/nature13460} {\bibfield  {journal} {\bibinfo  {journal} {Nature}\ }\textbf {\bibinfo {volume} {510}},\ \bibinfo {pages} {351–355} (\bibinfo {year} {2014})}\BibitemShut {NoStop}%
\bibitem [{\citenamefont {Howard}(2015)}]{Howard_2015}%
  \BibitemOpen
  \bibfield  {author} {\bibinfo {author} {\bibfnamefont {M.}~\bibnamefont {Howard}},\ }\href {https://doi.org/10.1103/PhysRevA.91.042103} {\bibfield  {journal} {\bibinfo  {journal} {Phys. Rev. A}\ }\textbf {\bibinfo {volume} {91}},\ \bibinfo {pages} {042103} (\bibinfo {year} {2015})}\BibitemShut {NoStop}%
\bibitem [{\citenamefont {Tirrito}\ \emph {et~al.}(2024)\citenamefont {Tirrito}, \citenamefont {Tarabunga}, \citenamefont {Lami}, \citenamefont {Chanda}, \citenamefont {Leone}, \citenamefont {Oliviero}, \citenamefont {Dalmonte}, \citenamefont {Collura},\ and\ \citenamefont {Hamma}}]{Tirrito_2024}%
  \BibitemOpen
  \bibfield  {author} {\bibinfo {author} {\bibfnamefont {E.}~\bibnamefont {Tirrito}}, \bibinfo {author} {\bibfnamefont {P.~S.}\ \bibnamefont {Tarabunga}}, \bibinfo {author} {\bibfnamefont {G.}~\bibnamefont {Lami}}, \bibinfo {author} {\bibfnamefont {T.}~\bibnamefont {Chanda}}, \bibinfo {author} {\bibfnamefont {L.}~\bibnamefont {Leone}}, \bibinfo {author} {\bibfnamefont {S.~F.~E.}\ \bibnamefont {Oliviero}}, \bibinfo {author} {\bibfnamefont {M.}~\bibnamefont {Dalmonte}}, \bibinfo {author} {\bibfnamefont {M.}~\bibnamefont {Collura}},\ and\ \bibinfo {author} {\bibfnamefont {A.}~\bibnamefont {Hamma}},\ }\href {https://doi.org/10.1103/PhysRevA.109.L040401} {\bibfield  {journal} {\bibinfo  {journal} {Phys. Rev. A}\ }\textbf {\bibinfo {volume} {109}},\ \bibinfo {pages} {L040401} (\bibinfo {year} {2024})}\BibitemShut {NoStop}%
\bibitem [{\citenamefont {Bejan}\ \emph {et~al.}(2024)\citenamefont {Bejan}, \citenamefont {McLauchlan},\ and\ \citenamefont {B\'eri}}]{Bejan_2024}%
  \BibitemOpen
  \bibfield  {author} {\bibinfo {author} {\bibfnamefont {M.}~\bibnamefont {Bejan}}, \bibinfo {author} {\bibfnamefont {C.}~\bibnamefont {McLauchlan}},\ and\ \bibinfo {author} {\bibfnamefont {B.}~\bibnamefont {B\'eri}},\ }\href {https://doi.org/10.1103/PRXQuantum.5.030332} {\bibfield  {journal} {\bibinfo  {journal} {PRX Quantum}\ }\textbf {\bibinfo {volume} {5}},\ \bibinfo {pages} {030332} (\bibinfo {year} {2024})}\BibitemShut {NoStop}%
\bibitem [{\citenamefont {Iannotti}\ \emph {et~al.}(2025)\citenamefont {Iannotti}, \citenamefont {Esposito}, \citenamefont {Campos~Venuti},\ and\ \citenamefont {Hamma}}]{Iannotti_2025}%
  \BibitemOpen
  \bibfield  {author} {\bibinfo {author} {\bibfnamefont {D.}~\bibnamefont {Iannotti}}, \bibinfo {author} {\bibfnamefont {G.}~\bibnamefont {Esposito}}, \bibinfo {author} {\bibfnamefont {L.}~\bibnamefont {Campos~Venuti}},\ and\ \bibinfo {author} {\bibfnamefont {A.}~\bibnamefont {Hamma}},\ }\href {https://doi.org/10.22331/q-2025-07-21-1797} {\bibfield  {journal} {\bibinfo  {journal} {{Quantum}}\ }\textbf {\bibinfo {volume} {9}},\ \bibinfo {pages} {1797} (\bibinfo {year} {2025})}\BibitemShut {NoStop}%
\bibitem [{\citenamefont {Gu}\ \emph {et~al.}(2025)\citenamefont {Gu}, \citenamefont {Oliviero},\ and\ \citenamefont {Leone}}]{Gu_2025}%
  \BibitemOpen
  \bibfield  {author} {\bibinfo {author} {\bibfnamefont {A.}~\bibnamefont {Gu}}, \bibinfo {author} {\bibfnamefont {S.~F.}\ \bibnamefont {Oliviero}},\ and\ \bibinfo {author} {\bibfnamefont {L.}~\bibnamefont {Leone}},\ }\href {https://doi.org/10.1103/PRXQuantum.6.020324} {\bibfield  {journal} {\bibinfo  {journal} {PRX Quantum}\ }\textbf {\bibinfo {volume} {6}},\ \bibinfo {pages} {020324} (\bibinfo {year} {2025})}\BibitemShut {NoStop}%
\bibitem [{\citenamefont {Mac\^edo}\ \emph {et~al.}(2025)\citenamefont {Mac\^edo}, \citenamefont {Andriolo}, \citenamefont {Zamora}, \citenamefont {Poderini},\ and\ \citenamefont {Chaves}}]{Macedo_2025}%
  \BibitemOpen
  \bibfield  {author} {\bibinfo {author} {\bibfnamefont {R.~A.}\ \bibnamefont {Mac\^edo}}, \bibinfo {author} {\bibfnamefont {P.}~\bibnamefont {Andriolo}}, \bibinfo {author} {\bibfnamefont {S.}~\bibnamefont {Zamora}}, \bibinfo {author} {\bibfnamefont {D.}~\bibnamefont {Poderini}},\ and\ \bibinfo {author} {\bibfnamefont {R.}~\bibnamefont {Chaves}},\ }\href {https://doi.org/10.1103/6srg-723m} {\bibfield  {journal} {\bibinfo  {journal} {Phys. Rev. A}\ }\textbf {\bibinfo {volume} {112}},\ \bibinfo {pages} {L050401} (\bibinfo {year} {2025})}\BibitemShut {NoStop}%
\bibitem [{\citenamefont {Cusumano}\ \emph {et~al.}(2025)\citenamefont {Cusumano}, \citenamefont {Venuti}, \citenamefont {Cepollaro}, \citenamefont {Esposito}, \citenamefont {Iannotti}, \citenamefont {Jasser}, \citenamefont {Odavi\'c}, \citenamefont {Viscardi},\ and\ \citenamefont {Hamma}}]{Cusumano_2025}%
  \BibitemOpen
  \bibfield  {author} {\bibinfo {author} {\bibfnamefont {S.}~\bibnamefont {Cusumano}}, \bibinfo {author} {\bibfnamefont {L.~C.}\ \bibnamefont {Venuti}}, \bibinfo {author} {\bibfnamefont {S.}~\bibnamefont {Cepollaro}}, \bibinfo {author} {\bibfnamefont {G.}~\bibnamefont {Esposito}}, \bibinfo {author} {\bibfnamefont {D.}~\bibnamefont {Iannotti}}, \bibinfo {author} {\bibfnamefont {B.}~\bibnamefont {Jasser}}, \bibinfo {author} {\bibfnamefont {J.}~\bibnamefont {Odavi\'c}}, \bibinfo {author} {\bibfnamefont {M.}~\bibnamefont {Viscardi}},\ and\ \bibinfo {author} {\bibfnamefont {A.}~\bibnamefont {Hamma}},\ }\href {https://arxiv.org/abs/2504.03351} {\bibinfo {title} {Non-stabilizerness and violations of chsh inequalities}} (\bibinfo {year} {2025}),\ \Eprint {https://arxiv.org/abs/2504.03351} {arXiv:2504.03351 [quant-ph]} \BibitemShut {NoStop}%
\bibitem [{\citenamefont {Cie\ifmmode \acute{s}\else \'{s}\fi{}li\ifmmode~\acute{n}\else \'{n}\fi{}ski}\ \emph {et~al.}(2026)\citenamefont {Cie\ifmmode \acute{s}\else \'{s}\fi{}li\ifmmode~\acute{n}\else \'{n}\fi{}ski}, \citenamefont {Knips}, \citenamefont {Weinfurter},\ and\ \citenamefont {Laskowski}}]{Cieslinski_2026}%
  \BibitemOpen
  \bibfield  {author} {\bibinfo {author} {\bibfnamefont {P.}~\bibnamefont {Cie\ifmmode \acute{s}\else \'{s}\fi{}li\ifmmode~\acute{n}\else \'{n}\fi{}ski}}, \bibinfo {author} {\bibfnamefont {L.}~\bibnamefont {Knips}}, \bibinfo {author} {\bibfnamefont {H.}~\bibnamefont {Weinfurter}},\ and\ \bibinfo {author} {\bibfnamefont {W.}~\bibnamefont {Laskowski}},\ }\href {https://doi.org/10.1103/1nkl-sphd} {\bibfield  {journal} {\bibinfo  {journal} {Phys. Rev. A}\ }\textbf {\bibinfo {volume} {113}},\ \bibinfo {pages} {052404} (\bibinfo {year} {2026})}\BibitemShut {NoStop}%
\bibitem [{\citenamefont {Viscardi}\ \emph {et~al.}(2026)\citenamefont {Viscardi}, \citenamefont {Dalmonte}, \citenamefont {Hamma},\ and\ \citenamefont {Tirrito}}]{Viscardi_2026}%
  \BibitemOpen
  \bibfield  {author} {\bibinfo {author} {\bibfnamefont {M.}~\bibnamefont {Viscardi}}, \bibinfo {author} {\bibfnamefont {M.}~\bibnamefont {Dalmonte}}, \bibinfo {author} {\bibfnamefont {A.}~\bibnamefont {Hamma}},\ and\ \bibinfo {author} {\bibfnamefont {E.}~\bibnamefont {Tirrito}},\ }\href {https://doi.org/10.21468/SciPostPhysCore.9.1.012} {\bibfield  {journal} {\bibinfo  {journal} {SciPost Phys. Core}\ }\textbf {\bibinfo {volume} {9}},\ \bibinfo {pages} {012} (\bibinfo {year} {2026})}\BibitemShut {NoStop}%
\bibitem [{\citenamefont {Fishman}\ \emph {et~al.}(2022)\citenamefont {Fishman}, \citenamefont {White},\ and\ \citenamefont {Stoudenmire}}]{ITensor}%
  \BibitemOpen
  \bibfield  {author} {\bibinfo {author} {\bibfnamefont {M.}~\bibnamefont {Fishman}}, \bibinfo {author} {\bibfnamefont {S.~R.}\ \bibnamefont {White}},\ and\ \bibinfo {author} {\bibfnamefont {E.~M.}\ \bibnamefont {Stoudenmire}},\ }\href {https://doi.org/10.21468/SciPostPhysCodeb.4} {\bibfield  {journal} {\bibinfo  {journal} {SciPost Phys. Codebases}\ ,\ \bibinfo {pages} {4}} (\bibinfo {year} {2022})}\BibitemShut {NoStop}%
\bibitem [{\citenamefont {Tran}\ \emph {et~al.}(2017)\citenamefont {Tran}, \citenamefont {Zuppardo}, \citenamefont {de~Rosier}, \citenamefont {Knips}, \citenamefont {Laskowski}, \citenamefont {Paterek},\ and\ \citenamefont {Weinfurter}}]{Tran_2017}%
  \BibitemOpen
  \bibfield  {author} {\bibinfo {author} {\bibfnamefont {M.~C.}\ \bibnamefont {Tran}}, \bibinfo {author} {\bibfnamefont {M.}~\bibnamefont {Zuppardo}}, \bibinfo {author} {\bibfnamefont {A.}~\bibnamefont {de~Rosier}}, \bibinfo {author} {\bibfnamefont {L.}~\bibnamefont {Knips}}, \bibinfo {author} {\bibfnamefont {W.}~\bibnamefont {Laskowski}}, \bibinfo {author} {\bibfnamefont {T.}~\bibnamefont {Paterek}},\ and\ \bibinfo {author} {\bibfnamefont {H.}~\bibnamefont {Weinfurter}},\ }\href {https://doi.org/10.1103/PhysRevA.95.062331} {\bibfield  {journal} {\bibinfo  {journal} {Phys. Rev. A}\ }\textbf {\bibinfo {volume} {95}},\ \bibinfo {pages} {062331} (\bibinfo {year} {2017})}\BibitemShut {NoStop}%
\bibitem [{\citenamefont {Kožić}\ and\ \citenamefont {Torre}(2025)}]{Kozic_2025}%
  \BibitemOpen
  \bibfield  {author} {\bibinfo {author} {\bibfnamefont {S.~B.}\ \bibnamefont {Kožić}}\ and\ \bibinfo {author} {\bibfnamefont {G.}~\bibnamefont {Torre}},\ }\href {https://arxiv.org/abs/2502.06956} {\bibinfo {title} {Computing quantum resources using tensor cross interpolation}} (\bibinfo {year} {2025}),\ \Eprint {https://arxiv.org/abs/2502.06956} {arXiv:2502.06956 [quant-ph]} \BibitemShut {NoStop}%
\end{thebibliography}%

\label{app:A}
\end{document}